%% file: paper.tex
\RequirePackage{etoolbox}
\documentclass[times,doublespace]{simauth}

\usepackage{amsmath}
\usepackage[colorlinks,citecolor=blue,urlcolor=blue,filecolor=blue,backref=page]{hyperref}
\usepackage{graphicx}
\usepackage{algorithm}
\usepackage{mathrsfs}
\usepackage{subcaption}
\usepackage{epstopdf}

\newcommand\BibTeX{{\rmfamily B\kern-.05em \textsc{i\kern-.025em b}\kern-.08em
T\kern-.1667em\lower.7ex\hbox{E}\kern-.125emX}}
\def\volumeyear{2010}

\newcommand{\bs}{\boldsymbol}

\begin{document}

\runninghead{Teng et al.}
\title{Time Series Analysis of fMRI Data: Spatial Modelling and Bayesian Computation}

\author{Ming Teng\affil{a}, Timothy D. Johnson\affil{a}, Farouk S. Nathoo\affil{b}\corrauth}
\address{\affilnum{a}  University of Michigan, Department of Biostatistics, 1415 Washington Heights, Ann Arbor, MI 48109 \\ 
\affilnum{b} University of Victoria, Department of Mathematics and Statistics, Victoria, BC, Canada, V8W 3P4}

\corraddr{University of Victoria, Department of Mathematics and Statistics, Victoria, BC, Canada, V8W 3P4. E-mail: nathoo@uvic.edu}

%\author{Ming Teng\affilnum{1},Timothy D. Johnson\affilnum{1}, Farouk S. Nathoo\affilnum{2}}
%\affiliation{\affilnum{1} University of Michigan \\
%\affilnum{2} University of Victoria}
%\corrauth{Farouk S. Nathoo, University of Victoria, Department of Mathematics and Statistics, Victoria, BC, Canada, V8W 3P4}
%\email {nathoo@uvic.edu}

\input{paper--abstract}

\maketitle

\input{paper--introduction}

\input{paper--methods}

\input{paper--results}

\input{paper--discussion}

 \begin{acks}
 F.S. Nathoo  is supported by funding from the Natural Sciences and Engineering Research Council of Canada and holds a Tier II Canada Research Chair in Biostatistics for Spatial and High-Dimensional Data. 
 \end{acks}

\bibliographystyle{wileyj}
\bibliography{paper}

\input{paper--tables}

\end{document}

%% file: paper--abstract.tex
\begin{abstract}
Time series analysis of fMRI data is an important area of medical statistics for neuroimaging data. Spatial models and Bayesian approaches for inference in such models have advantages over more traditional mass univariate approaches; however, a major challenge for such analyses is the required computation. As a result, the neuroimaging community has embraced approximate Bayesian inference based on mean-field variational Bayes (VB) approximations. These approximations are implemented in standard software packages such as 
the popular Statistical Parametric Mapping (SPM) software. While computationally efficient, the quality of VB approximations remains unclear even though they are commonly used in the analysis of neuroimaging data. For reliable statistical inference, it is important that these approximations be accurate and that users understand the scenarios under which they may not be accurate. 

We consider this issue for a particular model that includes spatially-varying coefficients. To examine the accuracy of the VB approximation we derive Hamiltonian Monte Carlo (HMC) for this model and conduct simulation studies to compare its performance with VB in terms of estimation accuracy, posterior variability, the spatial smoothness of estimated images, and computation time. As expected we find that the computation time required for VB is considerably less than that for HMC. In settings involving a high or moderate signal-to-noise ratio (SNR) we find that the two approaches produce very similar results suggesting that the VB approximation is useful in this setting. On the other hand, when one considers a low SNR, substantial differences are found, suggesting that the approximation may not be accurate in such cases and we demonstrate that VB produces Bayes estimators with 
larger mean squared error (MSE). A comparison of the two computational approaches in an application examining the haemodynamic response to face perception in addition to a comparison with the traditional mass univariate approach in this application is also considered. Overall, our work clarifies the usefulness of VB for the spatiotemporal analysis of fMRI data, while also pointing out the limitation of VB when the SNR is low and the utility of HMC in this case. 
\end{abstract}

\keywords{Hamiltonian Monte Carlo; Variational Bayes; fMRI; SPM; spatial model; time series}

%% file: paper--introduction.tex
\section{Introduction}
It is well known that fMRI data exhibit both spatial and temporal autocorrelation. A widely used approach for the analysis of such data is the general linear model with autoregressive errors and spatial smoothing priors for the regression coefficients (GLM-AR). Models of this sort have been developed in the Bayesian framework  (\cite{penny2005bayesian}; \cite{penny2007bayesian}; \cite{nathoo2013skew}) with approximate Bayesian inference based on mean field variational Bayes (VB). The VB approximation is used to handle the very large parameter space across voxels in the brain while maintaining computational tractability. While this approach often leads to computational efficiency, there are potential concerns with its accuracy. \cite{nathoo2013comparing} have discussed this issue and demonstrated examples with neuroimaging data where the mean field variational Bayes approximation can severely underestimate posterior variability and produce biased estimates of model hyper-parameters. \cite{yu2016understanding} study the performance of VB in a simulation study based on fMRI and raise concerns that while VB reduces computational cost it can suffer from lower specificity and smaller coverage of the credible intervals.

Simulation-based approaches for Bayesian computation such as importance sampling and Markov chain Monte Carlo (MCMC) have an underlying large sample theory guaranteeing simulation-consistent approximation (\cite{robert2013monte}) of various aspects of the posterior distribution, such as posterior moments and quantiles. Unfortunately, there is currently no such theory guaranteeing or characterizing the accuracy for VB approximations. This is an important issue as commonly used software packages such as SPM implement Bayesian approaches using VB without much consideration for its accuracy. As a result these approximations need to be checked on a case-by-case basis, typically against the output from properly tuned MCMC algorithms. In some cases, the quality of the VB approximation will be very good and in other cases the VB approximation can be quite poor. For a given model where the VB approximation is used, it is of practical importance for users to have some general understanding of the quality of this approximation, and if computational resources are available, to be able to check this for certain test cases (e.g. using the fMRI data from a select few subjects in a study). 

In making comparisons with MCMC algorithms, it is important that the particular MCMC algorithm being employed achieves adequate mixing and thus is able to traverse the parameter space fairly rapidly. This is a particularly important issue when dealing with spatial models for fMRI data as the number of parameters in the model and their potentially high posterior correlations can result in poor performance of standard MCMC algorithms such as the Gibbs sampler and the random walk Metropolis-Hastings algorithm, as well as algorithms that combine Gibbs and random walk Metropolis-Hastings moves. MCMC algorithms of this sort for spatio-temporal fMRI time series models have been developed by \cite{woolrich2004fully} where Gibbs sampling and single-component Metropolis-Hastings jumps are employed for posterior simulation. An alternative MCMC algorithm that is better suited for large parameter spaces with high posterior correlations is the HMC algorithm (\cite{duane1987hybrid}; \cite{neal1995bayesian}). For neuroimaging data and dynamic causal modeling, the HMC algorithm has been recently explored by \cite{sengupta2016gradient} where it is found that HMC and Langevin Monte Carlo are far superior to the random walk Metropolis algorithm when applied for the estimation of neural mass models.  The derivation of HMC and its comparison to mean-field VB for the time series analysis of fMRI data with spatial priors has not been considered previously. As a practical matter, it is important for those who analyze neuroimaging data on a routine basis to understand the potential drawbacks of the approximations implemented in standard software such as SPM.

The primary goals of this paper are two-fold. First, considering autoregressive models for fMRI data with spatially-varying  (regression and autoregressive) parameters, we derive for the first time an HMC algorithm for implementing Bayesian inference. Second, we make detailed comparisons between our HMC algorithm and the VB algorithm currently used in the SPM software. Our goal is to determine when the VB approximation is accurate and when it is not, so as to provide practitioners with guidance for the best implementation of these models. Finally, we provide software for our HMC implementation as C code which can be obtained at the following link \url{http://www.math.uvic.ca/~nathoo/publications.html}.

In Section 2 we present the spatial fMRI model and the VB algorithm used for approximating the posterior distribution. We then derive HMC for this model and discuss the tuning of this algorithm. In Section 3 we present three simulation studies as well as a comparison of methods on a real fMRI dataset examining the neural response to face repetition \cite{henson2002face}. Section 4 concludes with a brief discussion.

%% file: paper--methods.tex
\section{Methods}
We begin by briefly discussing the fMRI spatial model. We then describe the variational Bayes (VB) and Hamiltonian Monte Carlo (HMC) algorithms that can be used to fit this model. We put a greater emphasis on the HMC algorithm as the VB algorithm has been discussed in \cite{penny2005bayesian}. 

\subsection{The fMRI spatial model}
We let $T$ denote the length of each time series, $N$ the number of voxels, $K$ the number of regressors in the linear model, and $P$ the order of the temporal autoregressive process used to model the temporal correlation at each voxel. Throughout this paper, a matrix is indicated with bold capital letters, while vectors are indicated with bold lower-case letters, and scalars are denoted by lower-case letters. The linear model at the $n^{th}$ voxel, n=1,\dots,N, is specified as
\begin{equation}
\mathbf y_{P+1:T,n}  = \mathbf X \mathbf w_n + \mathbf e_n 
\end{equation}
where $\mathbf y_n=(y_{1n},...,y_{Tn})^T$ denotes the time series of length $T$ recorded at the $n^{th}$ voxel with last $(T-P)$ components denoted as $\mathbf y_{P+1:T,n}$, and where we condition on the first $P$ components $\mathbf y_{1:P,n}$ for simplicity in implementation of the autoregressive model. $\mathbf X=(\mathbf x_1, ...,\mathbf x_K)$ denotes the $K$ columns of regressors each having length $T-P$; $\mathbf w_n$ is the corresponding vector of regression coefficients specific to voxel $n$. The regressors are typically stimulus indicators convolved with the haemodynamic response function (HRF), $x_{tk}=(v_k*h)(t)$, that is, the $k^{th}$ regressor at time $t$, is the $k^{th}$ stimulus $v_k$ convolved with the HRF $h(\cdot)$ at time $t$. Details are described in \cite{lindquist2008statistical}. The autoregressive process for the model errors is specified as
\begin{equation}
\mathbf e_n = \mathbf{\tilde E}_n \mathbf a_n + \mathbf z_n
\end{equation}
where $\mathbf{\tilde E}_n=(\mathbf{\tilde e}_{P+1,n},...,\mathbf{\tilde e}_{Tn})^T$ is a $(T-P) \times P$ lagged prediction matrix with $t^{th}$ row $\mathbf{ \tilde e}_{tn} =(e_{t-1,n},...,e_{t-P,n})$; $\mathbf a_n=(a_{1n},...,a_{pn})^T$ is the corresponding vector of auto-regressive coefficients for voxel n; $\mathbf z_n=(z_{P+1,n},...,z_{Tn})^T$ is Gaussian noise for voxel n, with $z_{tn}$ i.i.d with mean 0 and precision $\lambda_n \ (t=P+1,...T)$. The contribution to the log-likelihood for voxel n, is then:
\begin{equation} \label{eq1}
l_n=- \frac{\lambda_n}{2} \sum_{t=P+1}^T \left [  (y_{tn}- \mathbf x_t \mathbf w_n)- \mathbf{\tilde e}_{tn} \mathbf a_n    \right ]^2 + \frac{T-P}{2} \log \lambda_n +const 
\end{equation}
where $const$ denotes a constant that does not depend on the model parameters, and $\mathbf x_t$ is the $t^{th}$ row of $\mathbf X$. We note that this formulation conditions on the data observed at the first $P$ time points. It is in fact not necessary to condition on the first $P$ values of the time series. Without conditioning, a missing data approach (e.g. EM) can be applied rather than conditioning; however, conditioning simplifies the implementation and this simpler approach yields essentially equivalent results when $T$ is large compared with $P$. Thus the conditioning will typically have little effect on the resulting inference (\cite{penny2003variational}). The overall log-likelihood is then obtained by summing  $l_n$ across all voxels $l=\sum_{n=1}^N l_n$.

Regarding priors for the model parameters, let $\mathbf W=(\mathbf w_1,...,\mathbf w_N)$ denote the set of regression coefficients across all of the voxels, so that $\mathbf W$ is $K \times N$. The rows of $\mathbf W$ are assumed a priori independent, but the model adopts a prior that incorporates spatial dependence across voxels (across the columns of $\mathbf W$ within each row). This spatial dependence is practically important as it allows for smoothing across voxels and the borrowing of information spatially. Let $\mathbf w_k$ be the $k^{th}$ row of $\mathbf W$, a vector of length $N$, and let $\pi(\mathbf W|\bs\alpha)$ denote the prior density which takes the form
\begin{eqnarray} \label{eqwprior}
&& \pi(\mathbf W|\alpha)= \prod_{k=1}^K \pi(\mathbf w_k^T|\alpha_k) \nonumber \\
&& \mathbf w_k^T \mid \alpha_k \sim \mbox N(\mathbf 0, \alpha_k^{-1} (\mathbf S^T \mathbf S )^{-1}).
\end{eqnarray}
where $\bs\alpha=(\alpha_1,...,\alpha_K)^T$ are hyper-parameters. Here $\mathbf S$ is a spatial kernel and takes the form of a non-singular Laplacian matrix (\cite{pascual1994low}) with elements:

\begin{equation}
 s_{ij}=
    \begin{cases}
      deg, & \text{if}  \ i=j \\
      -1, & \text{if}\ i \neq j \ \text{and i is adjacent to j} \\
      0, & \text{otherwise} 
    \end{cases}
\end{equation}
where $deg=4$ for a two dimensional model and $deg=6$ for a three dimensional model. These choices for the degree of the spatial kernel follow from \cite{penny2005bayesian}. By formulating the spatial kernel matrix in this way, smoothing is achieved and it is easy to show that the precision matrix $\mathbf S^T \mathbf S$ is a sparse matrix with 13 non-zero elements on each row and each column for a two dimensional model, and 25 non-zero elements on each row and each column for a three dimensional model. 

In addition to assigning spatial priors to the regression coefficients the model also contains temporal autoregressive parameters at each voxel and these are also assigned spatial priors allowing for spatial smoothing in the temporal autocorrelation parameters across the brain. Let $\mathbf A=(\mathbf a_1,...,\mathbf a_N)$ denote the autoregressive parameters across all voxels and let $\mathbf a_p$ denote the $p^{th}$ row of $\mathbf A$, the prior for $\mathbf A$ is
\begin{eqnarray}\label{eqaprior}
&& \pi(\mathbf A|\bs\beta)=\prod_{p=1}^P \pi(\mathbf a_p^T|\beta_p) \nonumber \\
&& \mathbf a_p^T \mid \beta_p \sim \mbox N(\mathbf 0,\beta_p^{-1}(\mathbf D^T \mathbf D)^{-1})
\end{eqnarray}
where $\bs\beta=(\beta_1,...,\beta_P)^T$ are hyper-parameters; $\mathbf D$ is a spatial kernel matrix similar to $\mathbf S$, for simplicity we will assume that $\mathbf D=\mathbf S$. 

For the hyper-parameters $\bs\alpha=(\alpha_1,...,\alpha_K)^T$, $\bs\beta=(\beta_1,...,\beta_P)^T$, and precision parameters $\bs\lambda=(\lambda_1,...,\lambda_N)^T$, the model assumes that these parameters are conditionally independent with each following a Gamma distribution a priori:
\begin{eqnarray}
&& \pi(\bs\alpha \mid q_1, q_2)=\prod_{k=1}^K \pi(\alpha_k \mid q_1, q_2) \\
&& \alpha_k \mid q_1, q_2 \sim  G(q1,q2) \\
&& \pi(\bs\beta \mid r_1, r_2)=\prod_{p=1}^P \pi(\beta_p \mid r_1, r_2) \\
&& \beta_p \mid r_1, r_2 \sim  G(r1,r2) \\
&& \pi(\bs\lambda \mid u_1, u_2)=\prod_{n=1}^N \pi(\lambda_n \mid u_1, u_2) \\
&& \lambda_n \mid u_1, u_2 \sim G(u1,u2) 
\end{eqnarray}
where $G(q_1,q_2)$ denotes the density of the Gamma distribution with mean $q_1q_2$ and variance $q_1q_2^2$ and $q_1,q_2,r_1,r_2,u_1,u_2$ are fixed known values. In what follows we assume that $q_1=r_1=u_1=0.01$ and $q_2=r_2=u_2=100$.

Let $\bs\theta=(\mathbf w_1,...,\mathbf w_K,\mathbf a_1,...,\mathbf a_P,\bs\alpha^T,\bs\beta^T,\bs\lambda^T)^T$ denote the set of all parameters stacked in row-major order, we have $\text{dim}(\bs\theta) = R$ where $R=(K+P+1)N+K+P$,  and the log of the posterior density is
\begin{eqnarray}\label{posterior}
&&  \log p(\bs\theta \mid \mathbf Y, \mathbf X)
 =\sum_{n=1}^N \left \{ - \frac{\lambda_n}{2} \sum_{t=P+1}^T \left [  (y_{tn}- \mathbf x_t \mathbf w_n)- \mathbf{\tilde e}_{tn} \mathbf a_n    \right ]^2 \right \}  \nonumber \\
&& + \frac{T-P}{2} \sum_{n=1}^N \log \lambda_n
+ \sum_{k=1}^K \left [ -\frac{1}{2} \mathbf w_k (\alpha_k ( \mathbf S^T S) ) \mathbf w_k^T + \frac{1}{2} \log (| \alpha_k (\mathbf S^T \mathbf S) | ) \right ] \nonumber \\
&&  +\sum_{p=1}^P \left [ -\frac{1}{2} \mathbf a_p (\beta_p (\mathbf D^T \mathbf D)) \mathbf a_p^T + \frac{1}{2} \log | \beta_p (\mathbf D^T \mathbf D) |  \right ]+\sum_{k=1}^K \left [ (q_1-1) \log \alpha_k - \alpha_k / q_2  \right ] 
  \nonumber \\
 && +  \sum_{p=1}^P \left [ (r_1-1) \log \beta_p - \beta_p / r_2     \right ]  
 + \sum_{n=1}^N \left [ (u_1-1) \log \lambda_n -\lambda_n/ u_2 \right ] +const
 \end{eqnarray}
 where $\mathbf Y=(\mathbf y_1,...,\mathbf y_N)$ is the fMRI response data. Bayesian inference for the various components of $\bs\theta$ requires computation of the corresponding appropriately normalized posterior marginal distributions. Strategies for this Bayesian computation are described in what follows.

\subsection{Algorithm A: Variational Bayes}

Variational Bayes is an optimization approach for constructing a deterministic approximation to the posterior distribution. Let $q(\bs\theta)$ be a density function having the same support as the posterior density $p(\bs\theta \mid \mathbf Y, \mathbf X)$, and let $\log p(\mathbf Y|\mathbf X)$ denote the logarithm of the marginal likelihood associated with the model and the response $\mathbf Y$, which depends on the known design $\mathbf X$. We can express the logarithm of the marginal likelihood as
\begin{eqnarray*}
\log p(\mathbf Y | \mathbf X ) &=& \int q(\bs\theta) \log \left\{  \frac{p(\mathbf Y, \bs\theta|\mathbf X)}{q(\bs\theta)}    \right\}d\bs\theta \\ &&+ \int q(\bs\theta)\log \left\{ \frac{q(\bs\theta)}{p(\bs\theta \mid \mathbf Y, \mathbf X)} \right \}d\bs\theta \\
&\ge& \int q(\bs\theta) \log \left\{  \frac{p(\mathbf Y, \bs\theta|\mathbf X)}{q(\bs\theta)}    \right\}d\bs\theta \equiv F(q) 
\end{eqnarray*}
such that the functional $F(q)$ is a lower bound for $\log p(\mathbf Y | \mathbf X )$ for any $q$. The approximation is obtained by restricting $q$ to a manageable class of density functions, and maximizing $F$ over that class. In this case the class of density functions over which the optimization is carried out is characterized by densities that can be factored as follows:
\begin{equation}
q(\bs\theta)=\prod_{n=1}^N q(\mathbf w_n) \prod_{n=1}^N q(\mathbf a_n) \prod_{k=1}^K q(\alpha_k) \prod_{p=1}^P q(\beta_p) \prod_{n=1}^N q(\lambda_n).
\end{equation}
Let $\mbox E_{-q_i}[\cdot]$ denote the expectation under $q$ for all parameters excluding the $i^{th}$ parameter. A coordinate ascent algorithm is applied to locally maximize $F$ based on update steps of the form 
\begin{equation}
q(\theta_i) \propto \exp \mbox E_{-q_i} [\log p(\mathbf Y,\bs\theta|\mathbf X) ] 
\end{equation}
which are iterated to convergence. Details can be found in \cite{penny2003variational} and \cite{jordan1999introduction}. As mentioned in Section 1, the resulting approximate posterior distribution can be a very good approximation or conversely a very poor approximation of the true posterior density. While there are a number of factors that govern the quality of the approximation, as far as we are aware, there is currently no theory characterizing the error associated with mean-field VB. The only feasible approach is to compare the VB approximation with an appropriately implemented MCMC algorithm that has an associated large sample theory.

\subsection{Algorithm B: Hamiltonian Monte Carlo}

Hamiltonian Monte Carlo (HMC) has its origins with the work of \cite{alder1959studies} and \cite{duane1987hybrid} and was popularized in the statistical literature by \cite{neal1995bayesian}. It is a Metropolis-Hastings algorithm that can be used to sample high-dimensional target distributions far more efficiently than algorithms based on random walk proposals, where the proposals for HMC are based on Hamiltonian dynamics. The algorithm works by introducing a Hamiltonian $ H(\bs\theta,  \bs\xi)$ defined as the sum of potential energy $U( \bs\theta)$ and kinetic energy $K( \bs\xi)$, and the dynamics are written as follows:

\begin{eqnarray*}
\frac{d\theta_i}{dt} &=& \frac{\partial H(\bs\theta, \bs\xi)}{\partial \xi_i}=\frac{\partial K(\bs\xi)}{\partial \xi_i}\\
\frac{d\xi_i}{dt} &=& -\frac{\partial H(\bs\xi, \bs\theta)}{\partial  \theta_i}=-\frac{\partial U( \bs\theta)}{\partial \theta_i}.
\end{eqnarray*}

The continuous variable $t$ here denotes the time evolution of the dynamic system, $i \ (i=1,...,R)$ denotes the $i^{th}$ index of the corresponding random vector. $U( \bs\theta)=-p(\bs\theta \mid \mathbf Y, \mathbf X)$ is the negative log probability density function of the distribution for $\bs\theta$ that we wish to sample from, and $K( \bs\xi)$ is defined as $K( \bs\xi)=\bs\xi^T \mathbf M^{-1} \bs\xi/2$ where $\bs\xi$ is an auxiliary random vector having the same dimension as $\bs\theta$. Here $\mathbf M$ is referred to as the 'mass matrix' and is typically assumed diagonal. In practice this system is solved using numerical integration techniques (\cite{neal2011mcmc}), most commonly the leapfrog method. For fixed $\delta>0$ the method is comprised of the following updates:
\begin{eqnarray}
&& \bs\xi(t+\delta/2)=\bs\xi(t)-\delta/2 \frac{\partial U}{\partial \bs\theta}(\bs\theta(t)) \\
&& \bs\theta(t+\delta)=\bs\theta(t)+\delta \mathbf M^{-1} \bs\xi(t+\delta/2) \\
&& \bs\xi(t+\delta/2+\delta_l)=\bs\xi(t+\delta/2)-(\delta_l)\frac{\partial U}{\partial \bs\theta}(\bs\theta(t+\delta)).
\end{eqnarray}
where the method starts with a so-called half-step in equation 16, then iterates equations 17 and 18 $L$ times, where $\delta_l=\delta$ when $l<L$ and $\delta_l=\delta/2$ when $l=L$. The resulting approximate solution is used as a proposed value for the next state of the Markov chain in the Metropolis-Hastings (MH) algorithm.  

The algorithm requires repeated calculation of the unnormalized log-posterior density and its gradient. A fast way to calculate the log-likelihood components is thus crucial. Previous MCMC methods for models similar to the one considered here (e.g. \cite{woolrich2004fully}) compute the log-likelihood by directly summing across voxels $n$ and time points $t$. As a more efficient alternative we propose a calculation of the log-likelihood that can omit the summation across $t$.  Let $\mathbf a^*_n=(- \mathbf 1,\mathbf a_n^T)^T$, so $a^*_{pn}=a_{pn}$ if $p \geq 1$ and $a^*_{pn}=-1$ if $p=0$. The log-likelihood contribution for voxel $n$ can be expressed as:
\begin{eqnarray}\label{newlike}
l_n=- \frac{\lambda_n}{2}  \mathbf a^{*T}_n \mathbf F \mathbf a^*_n  +\frac{T-P}{2} \log \lambda_n + const.
\end{eqnarray}
where the specific form of $\mathbf F$ and its derivation is given in Appendix A of the Supplementary Material. Under this formulation, the sum across $t$ can be pre-computed rather than computed at every iteration of the algorithm. This changes the computational complexity of the likelihood evaluation from $O(TNKP)$ to $O(NK^{2}P^{2})$. Since $K\times P$ is typically smaller than $T$, this can make the computation faster, in our experience $10$ to $20$ times faster for datasets of the size considered in Section 3. Based on this form of the log-likelihood the gradient of the log-posterior density is derived as:
\begin{eqnarray}
&& \nabla w_{kn} \log p(\bs\theta \mid \mathbf Y, \mathbf X) = \lambda_n \mathbf a^*_n \mathbf G \mathbf a^{*T}_n - \alpha_k (\mathbf S^T \mathbf S)_n \mathbf w_k^T   \\
&& \nabla a_{pn} \log p(\bs\theta \mid \mathbf Y, \mathbf X) = \lambda_n  \mathbf f_p \mathbf a^*_n   -\beta_p (\mathbf D^T \mathbf D)_n \mathbf a_p^T       \\
&& \nabla \alpha_k \log p (\bs\theta \mid \mathbf Y, \mathbf X) = 
-\frac{1}{2} \mathbf w_k (\mathbf S^T \mathbf S) \mathbf w_k^T + 
(\frac{N}{2} + q_1-1)/ \alpha_k - \frac{1}{q_2} \\
&& \nabla \beta_{p} \log p (\bs\theta \mid \mathbf Y, \mathbf X) = -\frac{1}{2} \mathbf a_p (\mathbf D^T \mathbf D) \mathbf a_p^T + (\frac{N}{2}+r_1-1)/ \beta_p -\frac{1}{r_2} \\
&& \nabla \lambda_n \log p(\bs\theta \mid \mathbf Y, \mathbf X) =
-\frac{1}{2} \mathbf a^{*T}_n \mathbf F \mathbf a^*_n   +
\frac{(T-P)/2+u_1-1}{\lambda_n} - \frac{1}{u_2}
\end{eqnarray}
where $(\mathbf S^T \mathbf S)_n$ and $(\mathbf D^T \mathbf D)_n$ denotes the $n^{th}$ row of $\mathbf S^T \mathbf S$ and $\mathbf D^T \mathbf D$ respectively. Specific derivations including the form of $\mathbf G$ and $\mathbf f_p$ are given in Appendix B of the Supplementary Material.

There are a variety of block updating schemes that can be employed when updating the parameters in the HMC algorithm. For simplicity, we have tried various component-wise updates and have found that component-wise updates lead to very poor mixing of the sampling chain. On the other hand, updating the entire parameter vector $\bs\theta$ as a single high-dimensional block works well and produces adequate mixing when HMC is applied to this model. Letting $*$ indicate the current state of the sampling chain, the HMC algorithm proceeds as in Algorithm 1.
\begin{algorithm}[htbp]
\caption{HMC for GLM-AR}
\begin{enumerate}

\item Initialize the parameters $\bs\theta$, mass matrix $\mathbf M$, and Leapfrog step size $\delta$ and step number $L$.

\item  Update $\bs\theta$:\\

\begin{enumerate}
\item Simulate latent vector $\bs\xi^* \sim \mbox N(\mathbf 0, \mathbf I)$. Let $\bs\theta^{(0)} = \bs\theta^*$, $\bs\xi^{(0)}= \bs\xi^* + \frac{\delta}{2} \nabla_{\bs\theta} \log p( \bs\theta^*  \mid \mathbf Y)$

\item For $l=1,...,L$, let

\[
\bs\theta^{(l)} = \bs\theta^{(l-1)} + \delta / \mathbf M  \bs\xi^{(l-1)}
\]

\[
\bs\xi^{(l)} = \bs\xi^{(l-1)} + \delta^{(l)} \nabla \log p (\bs\theta^{(l)} \mid \mathbf Y, \mathbf X)
\]
where $\delta^{(l)}=\delta$ for $l <L $ and $\delta^{(L)} = \delta/2$

\item  Accept $\bs\theta^{(L)}$ as the new state for $\bs\theta$ with probability 
\[
\alpha_a = \min (1, e^{-H(\bs\theta^{(L)})+H(\bs\theta^*)}) 
\]
where $H(\bs\theta) = - \log p (\bs\theta \mid \mathbf Y, \mathbf X) + \bs\xi^T \mathbf M^{-1} \bs\xi /2$ \\
Else remain in the current state $\bs\theta^*$ with probability $1-\alpha_a$.
\\

\end{enumerate}
\item Repeat step 2 for the desired number of samples.
\end{enumerate}

\end{algorithm}

Tuning the HMC algorithm requires appropriate choice of $\mathbf M = \text{diag}\{m_1,\dots,m_R\}$, $\delta$, and $L$. We choose $\delta=0.00002$ as the initial value and adaptively adjust its value to obtain an optimal acceptance rate of around 0.65 (\cite{beskos2013optimal}) for a given value of $L$. Larger values for $L$ are useful in suppressing random walk behaviour of the chain, and we use $L=250$ in this work. Aside from examining the acceptance rate, mixing is judged from the output based on looking at the traceplots of some parameters specific to randomly chosen voxels, and we typically examine the traceplots of hyper-parameters as these components of the sampling chain often will mix slower than components corresponding to parameters higher up in the model hierarchy. Mixing is also judged based on estimation of the batch means Monte Carlo standard error (BMSE) (\cite{fishman1997implementation}), a measure that is easy to implement and is widely used in practice. 

As different parameters tend to have different scales, setting $m_i$ can also be important, and this issue is discussed extensively in \cite{neal2011mcmc}. In practice, we have found that for problems having moderate dimension and complexity, setting all $m_i=1 \ (i=1,...,R)$ is sufficient (e.g, Simulation 3.1). As the model complexity and dimensionality increases, we set the $m_i$ to be roughly proportional to the reciprocal of the posterior variance of the $i^{th}$ parameter for i=1,...,R.  This variance, of course, is unknown so it is estimated based on a preliminary run of HMC with $m_i=1 \ (i=1,...,R)$. This process is iterated a few times until adequate mixing of the chain is observed based on its output and the measures described above. We use this approach to tune the values of $\mathbf M$ in the application considered in Section 3.3.

%% file: paper--results.tex
\section{Results}

We conduct three simulation studies to compare features of the posterior distributions obtained from HMC and VB. The VB algorithm is implemented in the SPM12 software and for computations in this paper is run on MATLAB 2014a, on an iMac with 3.2 GHz and 16GB memory. The HMC algorithm code is written in C++, and implemented on the same machine in the case of our analysis of the face repetition data. For the simulation studies we run the HMC algorithm on a high-performance computing cluster (a Linux cluster powered by 12 dual quad-core Intel Xeon SMP compute nodes running at 2.33GHz per CPU). In all cases the HMC algorithm is run for 3000 iterations with first 2000 iterations discarded as burn-in, and the remaining 1000 iterations used to estimate features of the posterior distribution.

The simulation studies are followed by a real data analysis where we compare the results obtained from HMC, VB, and the traditional mass univariate approach. The simulation studies and application are based on the face-repetition dataset discussed in \cite{henson2002face}. A detailed description of this dataset can be found online at \emph{http://www.fil.ion.ucl.ac.uk/spm/data/}. The data are collected as part of an event-related fMRI study in which greyscale images of faces were repeatedly presented to a subject for 500 ms replacing the baseline, an oval chequerboard, that was present throughout the inter stimulus interval. Each of the faces were presented twice; some were familiar to the subject while others were not. This setup leads to four experimental conditions $U1$, $U2$, $F1$, $F2$, representing familiar or unfamiliar(F/U) faces observed for the first or second(1/2) time.

\begin{figure}[!ht]
\begin{center}
\begin{subfigure}{0.41\textwidth}
\includegraphics[width=\linewidth]{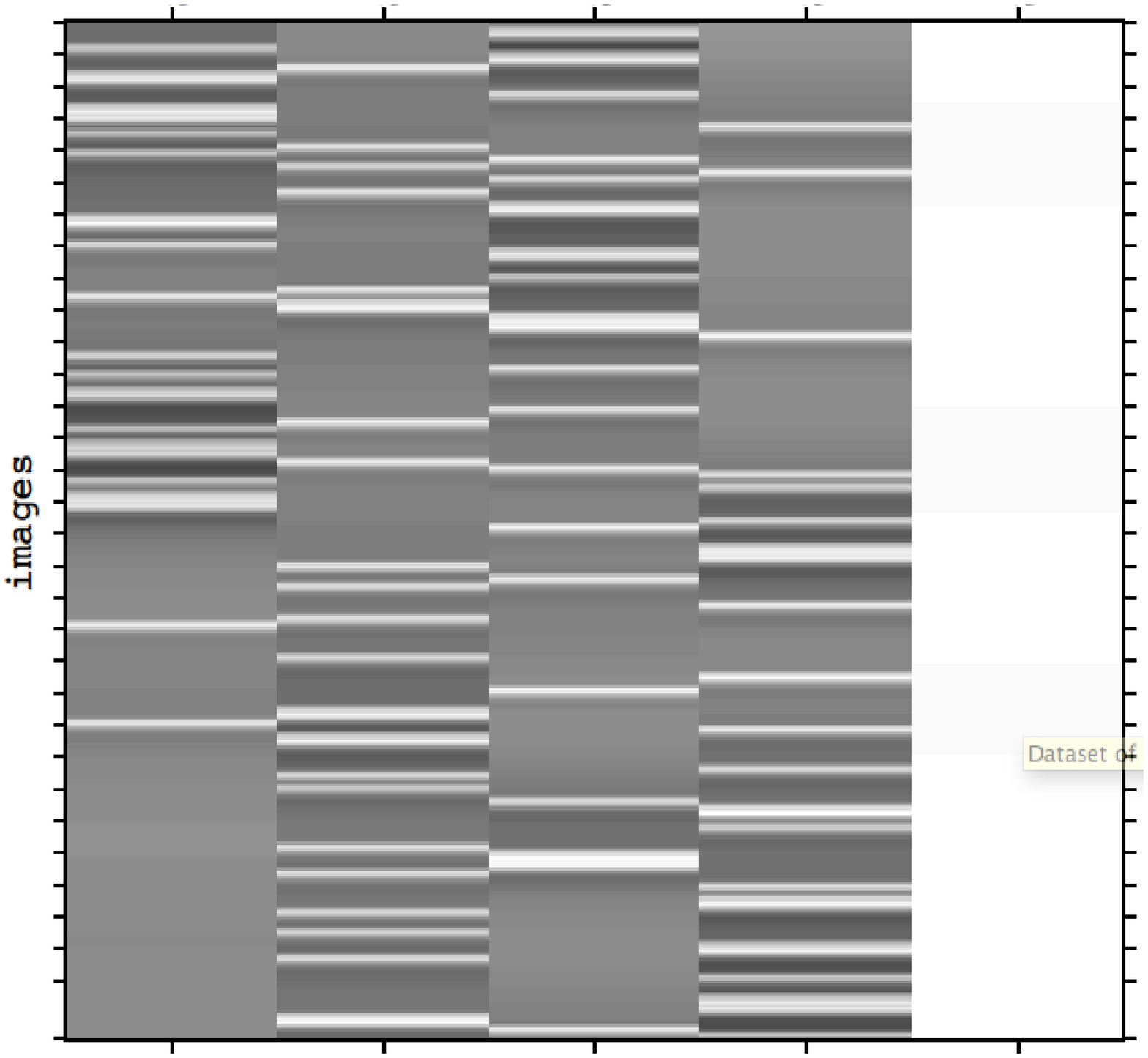}
\caption{} \label{fig_K5P1_matrix}
\end{subfigure}
\begin{subfigure}{0.41\textwidth}
\includegraphics[width=\linewidth]{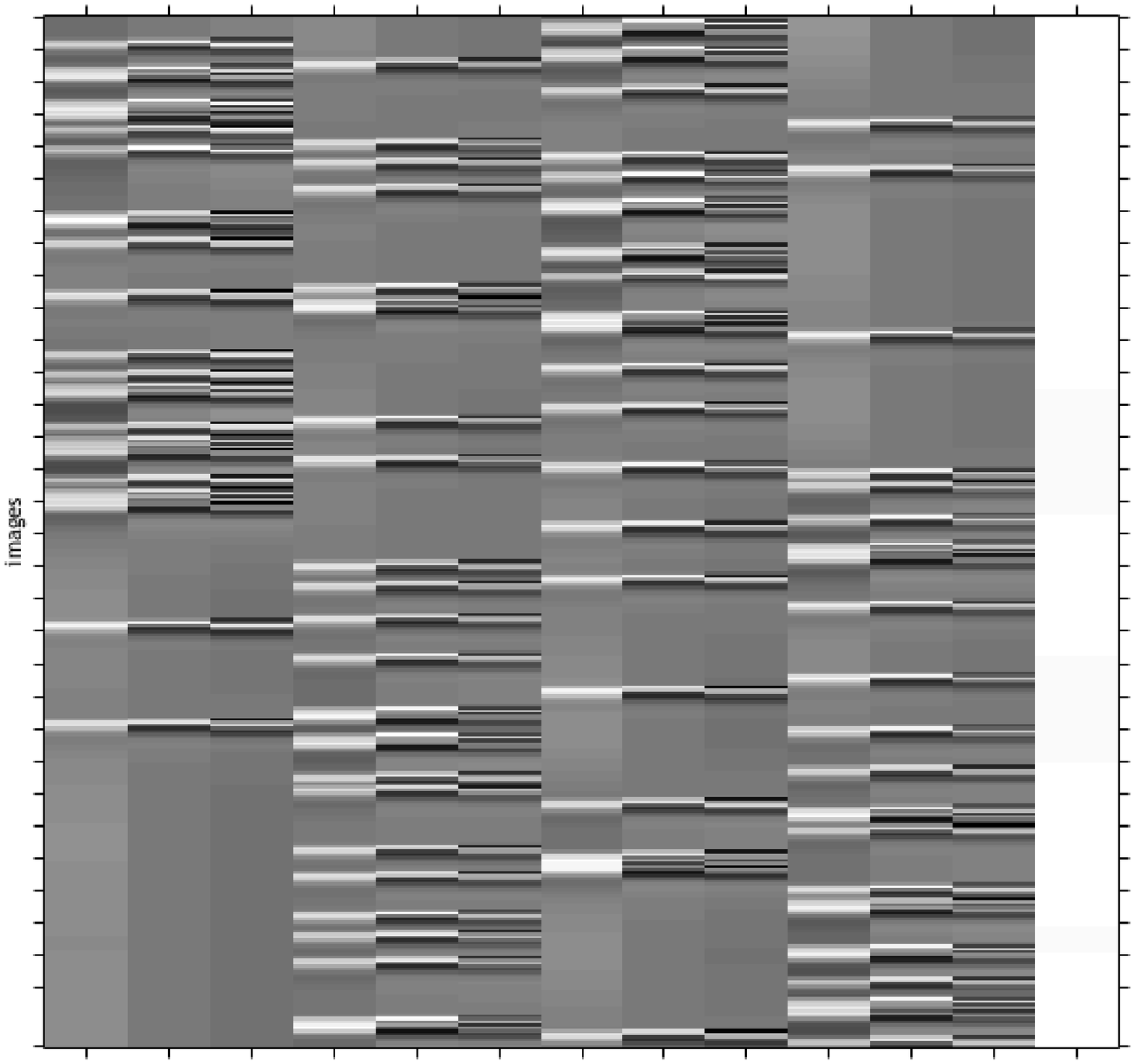}
\caption{} \label{fig_K13P3_matrix}
\end{subfigure}
\caption{\small{\textsf{Design matrix for simulation study one and three (a) and simulation study two (b). In panel (a), the first four columns correspond to stimuli U1, U2, F1, F2 convolved with the canonical HRF respectively. In panel (b), the 1st, 4th, 7th, and 10th columns are convolved with the canonical HRF, the 2nd, 5th, 8th, and 11th columns are convolved with its temporal derivative, the 3rd, 6th, 9th, and 12th columns are convolved with its dispersion derivative. The last blank column in both panels (a) and (b) represents the constant term.}}} 
\label{fig_matrix}
\end{center}
\end{figure}

The fMRI signal is measured at $T=351$ time points during the experiment. The design matrix used in the analysis has $(T-P)$ rows and $K$ columns. In our first and third simulation study we set $K=5$ corresponding to the four experimental conditions convolved with the canonical HRF, plus a constant term. The design matrix is depicted in Figure \ref{fig_K5P1_matrix}. In the second simulation study we consider a larger design matrix where each of the four study conditions is convolved with the canonical HRF, its dispersion derivative and its temporal derivative, respectively, resulting in $K=13$ columns (the last column corresponding to a constant term). The design matrix for the second simulation study is depicted in Figure \ref{fig_K13P3_matrix}.

For the simulation studies, we set the spatial domain to be a 2-dimensional lattice divided into a $53\times63$ grid, and then a brain-shaped mask is applied to this lattice, resulting in $N=2087$ voxels for the domain that our simulation studies are carried out on. The true values of the parameters $\mathbf W$, $\mathbf A$, and noise variables $\mathbf z_1,...,\mathbf z_N$ are simulated based on model assumptions and fixed values of $\bs\alpha$, $\bs\beta$, and $\bs\lambda$ discussed below. Given the parameter values, the data $\mathbf Y$ are simulated from the model and 100 replicate datasets are simulated in each study. To check consistency with respect to Monte Carlo variability, we have run all of the simulation studies with the same number of 100 replicate datasets twice and in each case the results from each of the two runs are consistent indicating stability. 

To compare VB and HMC with respect to point estimation, we use the simulation replicates and the known true values of the model parameters to estimate the average squared bias (ASBIAS) and the average mean squared error (AMSE) of estimators based on the posterior mean, where the average is taken across voxels. To compare the two approaches with respect to posterior variability we use the average marginal variance (AVAR). Letting $ \hat w_{knj} $ denote the posterior mean estimate of $w_{kn}$ obtained from the $j^{th} \ (j=1,...,J)$ simulation replicate, and $\sigma^2 ( \hat w_{knj})$ denote the corresponding posterior variance, the three measures above for $\mathbf w_k$  (where $k$ corresponds to the $k^{th}$ regressor) are computed as:
\begin{eqnarray}
&& \text{ASBIAS}(\mathbf w_k)=\frac{1}{N}\sum_{n=1}^N(\sum_{j=1}^J \hat w_{knj}/J-w_{kn})^2 \\
&& \text{AMSE} (\mathbf w_k)=\frac{1}{NJ} \sum_{n=1}^N \sum_{j=1}^J (\hat w_{knj}-w_{kn})^2 \\
&& \text{AVAR} (\mathbf w_k)=\frac{1}{NJ} \sum_{n=1}^N \sum_{j=1}^J \sigma^2 (\hat w_{knj})
\end{eqnarray}

These same measures are applied to the autoregressive coefficients $\mathbf a_p$. We also compute the correlation of each estimated $\mathbf w_k$ and $\mathbf a_p$ vectors with the truth, and average these correlations across simulation replicates.  To compare VB and HMC with respect to the spatial smoothness of the estimated images we use Moran's I (\cite{moran1950notes}). Negative values indicate negative spatial autocorrelation and positive values indicate positive spatial autocorrelation, a zero value corresponds to no spatial dependence. We compute Moran's I for each image of estimated parameters and then average these values (AMoran) across the J simulation replicates. For $\mathbf w_k$ this measure takes the form
\begin{equation}
\text{AMoran}=\frac{1}{J} \sum_{j=1}^J \frac{N}{\sum_{n_1} \sum_{n_2} \phi_{n_1n_2}} \frac{\sum_{n_1} \sum_{n_2} \phi_{n_1n_2}(w_{kn_1j}-\bar w_{kj})(w_{kn_2j}-\bar w_{kj})}{\sum_{n_1}(w_{kn_1j}-\bar w_{kj})^2}
\end{equation}
where $\bar w_{kj}= \sum_{n=1}^N w_{knj}$, $\phi_{n_1n_2}$ is the weight for voxel pair $(n_1,n_2) \ (n_1=1,...,N,n_2=1,...N)$, and here this is chosen as the reciprocal of the distance between the centroids of $n_1$ and $n_2$. 

%Simulation 1
\subsection{Simulation Study I}

We assume in this case that the data generating mechanism corresponds to a first-order autoregressive process. In simulating the true values of the regression coefficients and autoregressive coefficients we assign equal values to the precision of the regression coefficients, $\alpha_k=1 \ (k=1,...5)$ and we set $\beta_1=1000$ which will result in auto-regressive coefficients having much smaller values than the regression coefficients. For the precision of the noise we simulate these values from a Gamma distribution $\lambda_n \overset{\text{i.i.d}}{\sim} G(10,10) \ (n=1,...,N)$. 

\begin{figure}[!ht]
\begin{center}
\includegraphics[angle=-90,width=\linewidth]{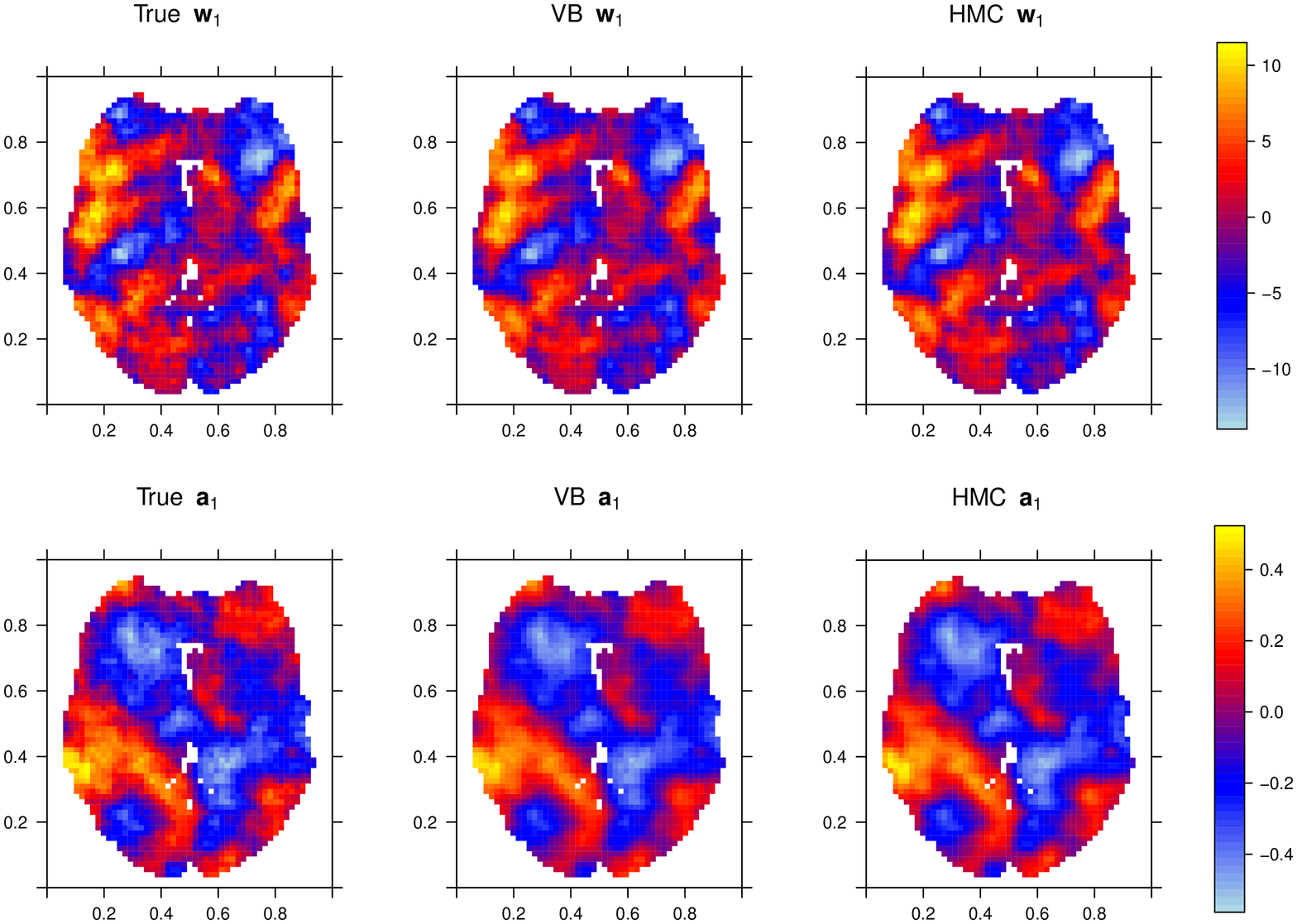}
\caption{\small{\textsf{Image of average (over simulation replicates) posterior mean estimate of $\mathbf w_1$ and $\mathbf a_1$ from HMC and VB. The estimates are compared with true image in each row. } }}
\label{fig_K5P1_image}
\end{center}
\end{figure}

Both VB and HMC are applied to the simulated datasets and images depicting the average (over simulation replicates) posterior mean estimates obtained from both methods and the true values are shown in Figure \ref{fig_K5P1_image}, where we show the images corresponding to $\mathbf w_1$ and $\mathbf a_1$. Figures depicting comparisons for the full set of parameters are shown in Figures 1-2 of the Supplementary Material. In this case the results obtained from HMC and VB are very similar and both correspond well with the truth.

The summary statistics discussed above are computed and their values are listed in Table \ref{tab_K5P1}. As the VB implementation in SPM does not provide the posterior variance of the auto-regressive coefficients as part of its output, we leave these cells blank in the table (including those for HMC since comparisons are of interest) . The statistics corresponding to HMC in the table are the actual values while those for VB are expressed as the percentage of the corresponding values obtained from HMC. From the table, we can see that VB  tends to produce smaller squared bias than HMC, but the MSE is roughly equivalent. The posterior variance statistics obtained from VB are also fairly close to those obtained from HMC, with slightly larger values for the former. Thus the over-confidence problem sometimes associated with VB (\cite{bishop2006pattern}, \cite{nathoo2013comparing}) does not seem to be an issue in this case. Both algorithms are performing well in terms of point estimation as they achieve a high level of correlation (around 0.99) with the true values. In terms of Moran's I, the images estimated using VB and HMC have approximately the same amount of spatial autocorrelation in their posterior estimates, and both are similar to the true Moran's I. In summary, VB and HMC both perform adequately well in this study.

%\begin{table}[!ht]
%\begin{center}
%\begin{tabular}{cccccccc}
%\hline \hline
%methods	&	measure	&	W1	&	W2	&	W3	&	W4	&	W5	&	A1	\\
%true	&	Moran's I	&	0.121	&	0.169	&	0.136	&	0.187	&	0.122	&	0.179	\\ \hline
%HMC	&	ASBIAS	&	0.123	&	0.120	&	0.105	&	0.110	&	0.001	&	4.52E-04	\\
%	&	AMSE	&	0.405	&	0.452	&	0.412	&	0.420	&	0.007	&	1.19E-03	\\
%	&	AVAR	&	0.411	&	0.468	&	0.425	&	0.435	&	0.008	&		\\
%	&	Correlation	&	0.997	&	0.999	&	0.998	&	0.998	&	1.000	&	0.995	\\
%	&	Moran's I	&	0.123	&	0.171	&	0.137	&	0.189	&	0.122	&	0.182	\\ \hline
%VB	&	ASBIAS	&	67\%	&	58\%	&	65\%	&	72\%	&	77\%	&	91\%	\\
%	&	AMSE	&	107\%	&	104\%	&	103\%	&	105\%	&	103\%	&	104\%	\\
%	&	AVAR	&	112\%	&	109\%	&	108\%	&	108\%	&	105\%	&		\\
%	&	Correlation	&	100\%	&	100\%	&	100\%	&	100\%	&	100\%	&	100\%	\\
%	&	Moran's I	&	100\%	&	100\%	&	100\%	&	100\%	&	100\%	&	102\%	\\
%\hline \hline
%\end{tabular}
%\end{center}
%\caption{\small{\textsf{Summary statistics for Simulation Study I. The results from VB are presented as a percentage of those obtained HMC. The true value of Moran's I is listed for each regressor in the first row as a reference.}}}
%\label{tab_K5P1}
%\end{table}
Comparing the two algorithms with respect to computation time on a standard iMac with 3.2 GHz Intel Core i5. HMC (coded in C++) takes 23min for 3000 iterations while VB takes 1min per simulated dataset. Overall, VB appears to perform well and result in an accurate approximation to the posterior distribution in this case. It also has the advantage of being computationally more efficient. 

%Simulation 2
\subsection{Simulation Study II}
In the second simulation study we aim to compare the performance of the two algorithms in a more complex situation, by including more regression coefficients at each voxel with these coefficients having unequal variance in the sense described below. Specifically, we extend the design matrix to include the canonical HRF, its temporal derivative, and its dispersion derivative. By convolving these functions with the four stimuli we get $13$ regressors (with the last corresponding to the constant term). We also increase the order of the auto-regressive process from $P=1$ to $P=3$. The precision parameters are set as follows: $\alpha_1=\alpha_2=\alpha_3=0.1$, $\alpha_4=\alpha_5=\alpha_6=0.5$, $\alpha_7=\alpha_8=\alpha_9=1.0$, $\alpha_{10}=\alpha_{11}=\alpha_{12}=2.0$, $\alpha_{13}=1.0$. $\beta_1=1000$, $\beta_2=2000$, $\beta_3=5000$. The values for the noise precision are again generated as $\lambda_{n} \overset{\text{i.i.d}}{\sim} Gamma(10,10) \ (n=1,...,N)$. 

\begin{figure}[!ht]
\begin{center}
\includegraphics[angle=-90,width=\linewidth]{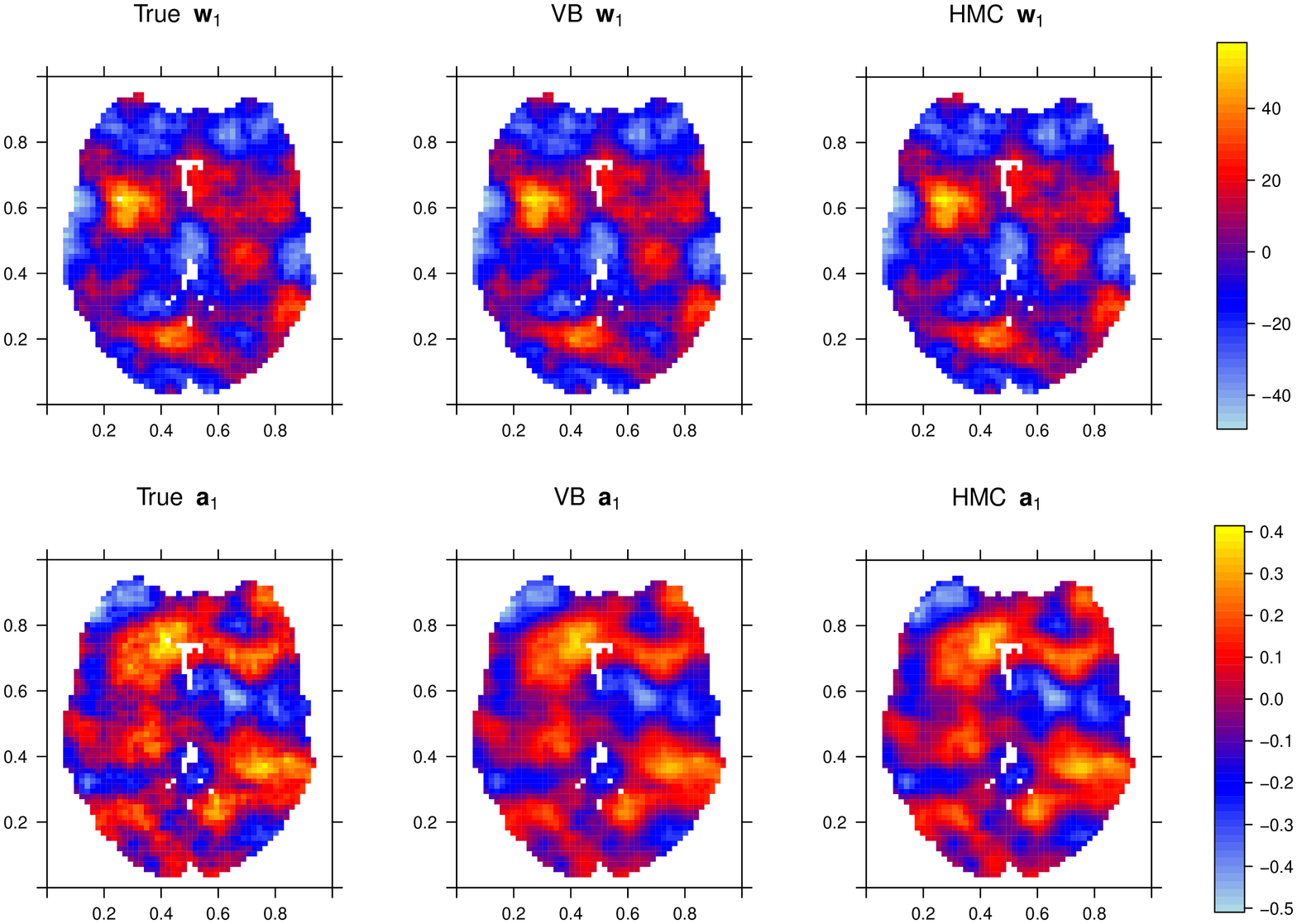}
\caption{\small{\textsf{Image of average (over simulation replicates) posterior mean estimate of $\mathbf w_1$ and $\mathbf a_1$ from HMC and VB. The estimates are compared with true image in each row.}}} 
\label{fig_K13P3_image}
\end{center}
\end{figure}

Figure \ref{fig_K13P3_image} shows the image of the average (over simulation replicates) posterior mean estimates from HMC and VB for $\mathbf w_1$ and $\mathbf a_1$. Similar Figures for the remaining parameters are shown in the Supplementary Material, Figures 3-8. Both HMC and VB appear to provide similar estimates which correspond well with the truth.

The summary statistics are computed as before and these are presented in Table \ref{tab_K13P3}. Generally, the observations made in Simulation Study I seem to carry over in that VB tends to produce smaller bias in point estimation but roughly equivalent MSE. Examining the average marginal posterior variance again indicates that VB does not exhibit an over-confidence problem in this case. The average correlation between the estimates and the truth obtained from HMC and VB are nearly the same, as seen in Study I. The measures of spatial correlation based on Moran's I are also again roughly equivalent for the two approaches.
In terms of timing, HMC takes 6.6 hours for 3000 iterations while VB takes 1 minute for a single simulation replicate. Overall, VB once again appears to perform well and results in an accurate approximation to the posterior distribution. In this second setting the gains in computational efficiency are rather substantial for VB. 

%Simulation 3
\subsection{Simulation Study III}
In the third simulation, we want to study the behavior of HMC and VB under a relatively low signal-to-noise ratio (SNR). We use the same design matrix and autoregressive order as in the first simulation study, where $K=5$ and $P=1$; however, instead of a high SNR, we lower the SNR by setting the precision of coefficients to be $\alpha_1=\alpha_2=\alpha_3=\alpha_4=100$, and $\beta_1=400$. The intercept has precision $\alpha_5=0.01$ and the precision for the noise term is set as $\lambda_n =0.1 (n=1,...,N)$. These settings correspond to a lower SNR compared with the first two studies and are therefore expected to pose a greater challenge in the estimation of the spatially-varying regression coefficients and autoregressive parameters.

\begin{figure}[!ht]
\begin{center}
\includegraphics[angle=-90,width=\linewidth]{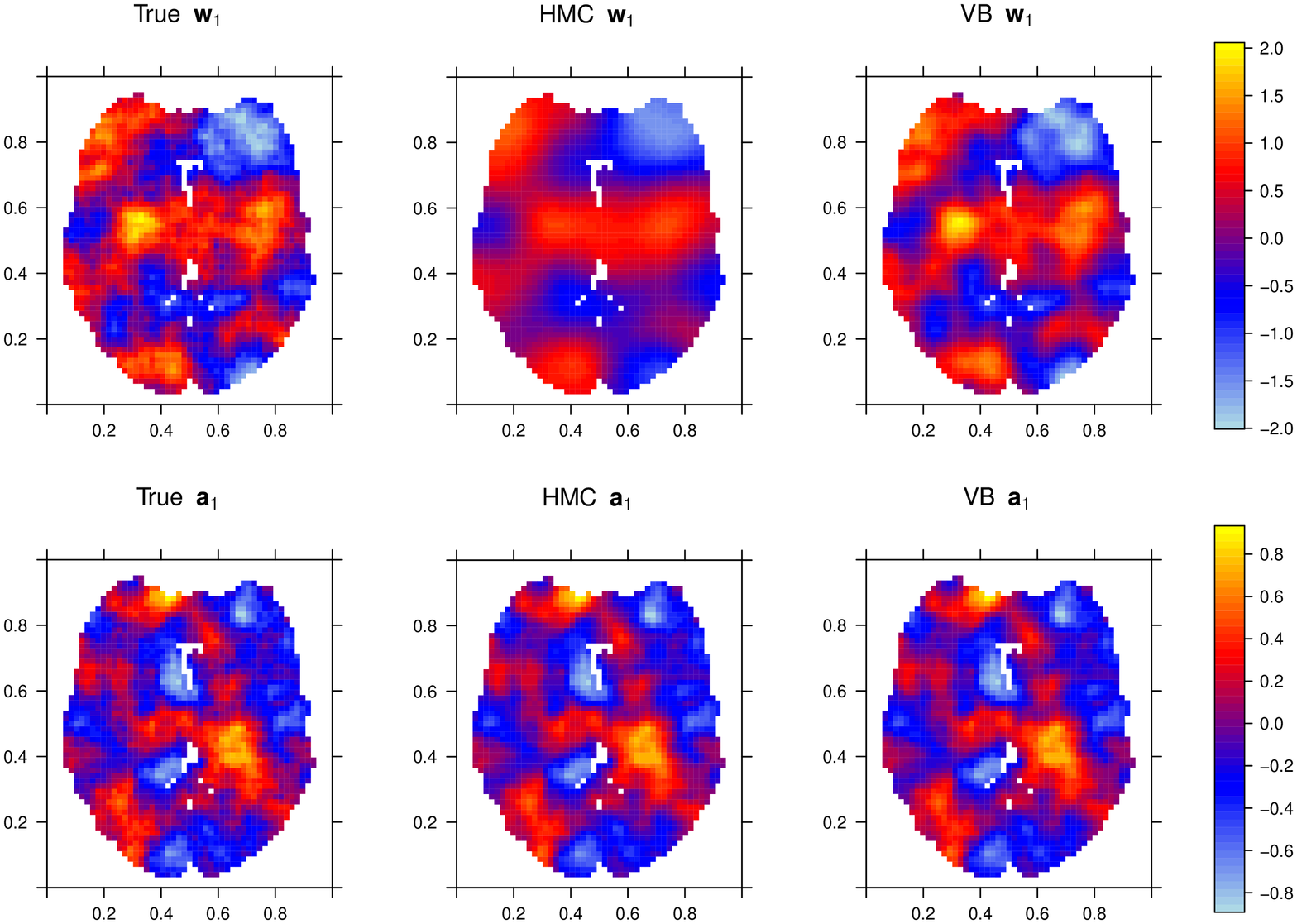}
\caption{\small{\textsf{Image of average (over simulation replicates) posterior mean estimate of $\mathbf w_1$ and $\mathbf a_1$ from HMC and VB, under a low SNR. The estimates are compared with true image in each row.}}} 
\label{fig_mean_SNR1e-3}
\end{center}
\end{figure}

Examining the results a difference is observed in the average (over simulation replicates) of the posterior mean, depicted in Figure \ref{fig_mean_SNR1e-3}. The estimates of the regression coefficients obtained from HMC appear to be smoother overall, and also tend to have larger bias when compared with those obtained from VB, while the two algorithms tend to give fairly similar results for the auto-regressive coefficients. A comparison of the numerical metrics is presented in Table \ref{tab_SNR1e-3}, and this comparison shows that VB actually over-estimates the posterior variability relative to HMC. The Bayes estimates obtained from VB actually have smaller bias, but the MSE obtained from HMC is considerably lower, in particular for the first four regression parameters. Here we see an example where VB results in a poor approximation to the posterior. Interestingly, the posterior variability is actually over-estimated rather than under-estimated as is typical. This demonstrates that when VB misses the mark it is not necessarily going to under-estimate posterior variability.

%\begin{table}[!ht]
%\begin{center}
%\begin{tabular}{cccccccc}
%\hline \hline
%methods	&	measure	&	W1	&	W2	&	W3	&	W4	&	W5	&	A1	\\
%true	&	Moran's I	&	0.167	&	0.203	&	0.202	&	0.197	&	0.147	&	0.122	\\ \hline
%HMC	&	ASBIAS	&	0.100	&	0.089	&	0.098	&	0.081	&	0.001	&	5.44E-04	\\
%	&	AMSE	&	0.163	&	0.168	&	0.175	&	0.165	&	0.046	&	1.56E-03	\\
%	&	AVAR	&	0.121	&	0.149	&	0.143	&	0.163	&	0.045	&		\\
%	&	Correlation	&	0.916	&	0.948	&	0.942	&	0.961	&	1.000	&	0.997	\\
%	&	Moran's I	&	0.201	&	0.260	&	0.193	&	0.209	&	0.147	&	0.124	\\ \hline
%VB	&	ASBIAS	&	19\%	&	18\%	&	19\%	&	21\%	&	61\%	&	83\%	\\
%	&	AMSE	&	416\%	&	437\%	&	370\%	&	410\%	&	110\%	&	101\%	\\
%	&	AVAR	&	441\%	&	391\%	&	352\%	&	328\%	&	99\%	&		\\
%	&	Correlation	&	107\%	&	104\%	&	105\%	&	103\%	&	100\%	&	100\%	\\
%	&	Moran's I	&	51\%	&	55\%	&	54\%	&	62\%	&	100\%	&	100\%	\\
%\hline \hline
%\end{tabular}
%\end{center}
%\caption{\small{\textsf{Summary statistics for Simulation Study I. The results from VB are presented as a percentage of those obtained HMC. The true value of Moran's I is listed for each regressor in the first row as a reference.}}}
%\label{tab_SNR1e-3}
%\end{table}

\begin{figure}[!ht]
\begin{center}
\includegraphics[angle=-90,width=\linewidth]{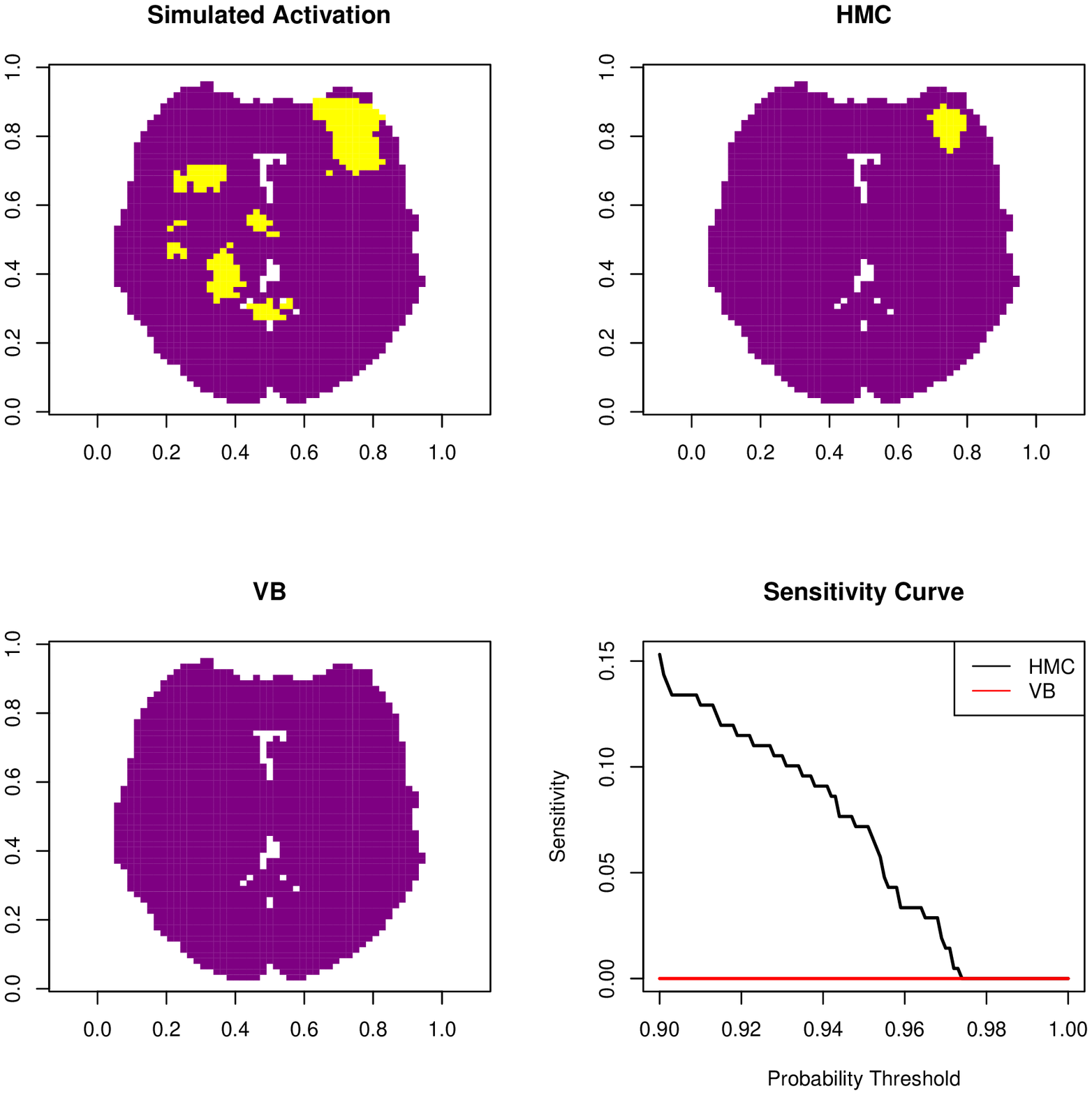}
\caption{\small{\textsf{Sensitivity curve and PPM showing the activated voxels. The PPM denotes simulated activation, activation inferred by HMC, and activation inferred by VB respectively. Blue dots denote the activation regions, while green denote no activation. The sensitivity curve is obtained by thresholding the probability from 0.9 to 1.}}} 
\label{fig_act_SNR1e-3}
\end{center}
\end{figure}

To quantify how much impact the observed difference will have on inference, we create a plot of sensitivity and posterior probability maps (PPM) on the effect of fame (one of the experimental conditions).  We do this by first defining a contrast vector $\mathbf c=(-1,-1,1,1, 0)^T/2$. We multiply this vector by $\mathbf w$, where $\mathbf w$ denotes the vector of regression coefficients at a given voxel, to get a contrast (or effect size) $\mathbf c^T \mathbf w$. We note that this contrast measures the effect of fame in the experiment at a given voxel. The posterior distribution of the contrast is then shown across voxels using a PPM. This map is based on two thresholds, the first being an effect size threshold $\gamma_e$ and the other being a probability threshold $\gamma_p$. The value of $\gamma_e$ is set to a constant so that top $10\%$ of the values across voxels of $\mathbf c^T \mathbf w$ are considered as "activated". The value of the probability threshold is set to be $\gamma_p=0.9$. At each voxel we then compute, using the posterior distribution,
\begin{equation}
Pr(\mathbf c^T \mathbf w > \gamma_e|\text{Data}) 
\end{equation}
and we highlight those voxels where the posterior probability is greater than $\gamma_p=0.9$. The PPM's obtained from HMC and VB, together with the sensitivity plot showing the true positives of fame, are depicted in Figure \ref{fig_act_SNR1e-3}. Examining this figure we note that there is a lack of power exhibited by both methods. Type 2 error is a problem in neuroimaging analysis in general but it is important to keep in mind that this is a difficult low SNR setting. From the figure, we see that HMC can capture about 15\% of the simulated activations at a probability threshold of $0.9$, while VB is unable to capture any of the activations. Overall, we see that in a low SNR setup the performance of VB is relatively poor when compared with HMC.

%%%%%%%%%%%%Applicaiton
\subsection{Real Application}
In this section, we will compare the estimation results from HMC, VB and the classical mass univariate approach (MUA) to examine possible differences in a real dataset obtained from a single subject. The dataset we focus on is again the face-repetition dataset; however, we now use the actual data and fit the model over the entire 3-dimensional brain volume based on a 3-dimensional grid having dimensions $53\times 63 \times52$ with a total of $56526$ voxels. Face recognition is also discussed in \cite{virji2012recognition}.

Pre-processing steps are conducted in SPM12: In order to remove the movement effects of the subject, all functional images are aligned to the first image using a six-parameter rigid-body transformation. To correct for differences in acquisition time of different slices, a slice timing correction procedure is also adopted. This procedure is done by first selecting the middle ($12^{th}$) slice as the reference slice, and then shifts the acquisition time of the remaining slices either forward or backward (depending on whether it's above or under the reference slice) to be at the same time point of the reference slice. Next, a coregistration procedure is performed by matching the mean functional image with the structural one based on a criteria that maximize their mutual information. Segmentation steps are then performed to create gray/white matter images, and bias-field corrected structural image. Images are also spatially normalized to a standard EPI template using a non-linear warping method. For MUA, the data are also pre-smoothed using a Gaussian kernel with FWHM of 8mm. This will help increase the signal to noise ratio, and also necessary for Gaussian assumption in classical framework to hold. We computed the global mean $g$ of all time series and scaled each time series by $100/g$; to further remove low frequency drift each time series was also high pass filtered using a default cutoff of $128s$. The design matrix is the same as that considered in Simulation Study I, shown in Figure \ref{fig_K5P1_matrix}. We fit the model with an autoregressive order of $P=1$ as in \cite{penny2005bayesian}. 

Both HMC and VB are initialized with starting values obtained from applying ordinary least squares regression (OLS) at each voxel. The hyper-parameters of the prior for the two algorithms are the same as those used previously, which corresponds to the default in the SPM software. For the mass matrix $\mathbf M$ in HMC, we use the tuning method described in Section 2.3. The trace plots for select parameters are displayed in the Supplementary Material, Figures 9-12, and these indicate adequate mixing of the sampling chain.

We note that the SPM implementation of VB when applied to analyze data over the whole brain volume uses a graph partitioning algorithm (\cite{harrison2008graph}). This works by dividing the whole brain into several disjoint regions and in each region the VB estimation is carried out independently. For this particular dataset, the graph partitioning algorithm divided the brain into $38$ regions.  Although this has the advantage of saving computational time, we find that this produces some artifacts as indicated below.

To compare the three methods with respect to point estimation we compute the correlation (across voxels) of the estimates, and these values are presented in Table \ref{tab_corr} which displays the correlation for each of the five regression coefficients $\mathbf w_1$ to $\mathbf w_5$ comparing VB and MUA to HMC. We see that HMC and VB have estimation (posterior mean) results that are highly correlated. The correlation between HMC and MUA for the intercept is only $0.66$; we suspect that pre-smoothing of the data (MUA) might be causing this relatively low value.

%\begin{table}[!ht]
%\begin{center}
%\begin{tabular}{ccccccc}
%\hline \hline
%											
%Correlation	&	$\mathbf w_1$	&	$\mathbf w_2$	&	$\mathbf w_3$	&	$\mathbf w_4$	&	$\mathbf w_5$	\\ \hline
%(VB, HMC)	&	0.91	&	0.92	&	0.91	&	0.89	&	1.00	\\
%(MUA, HMC)	&	0.87	&	0.84	&	0.84	&	0.83	&	0.66	\\
%\hline \hline
%\end{tabular}
%\end{center}
%\caption{\small{\textsf{Correlation (across voxels) in the estimated regression coefficients obtained from HMC and VB, and HMC and MUA.}}}
%\label{tab_corr}
%\end{table}

\begin{figure}[!ht]
\begin{center}
\includegraphics[angle=-90,width=\linewidth]{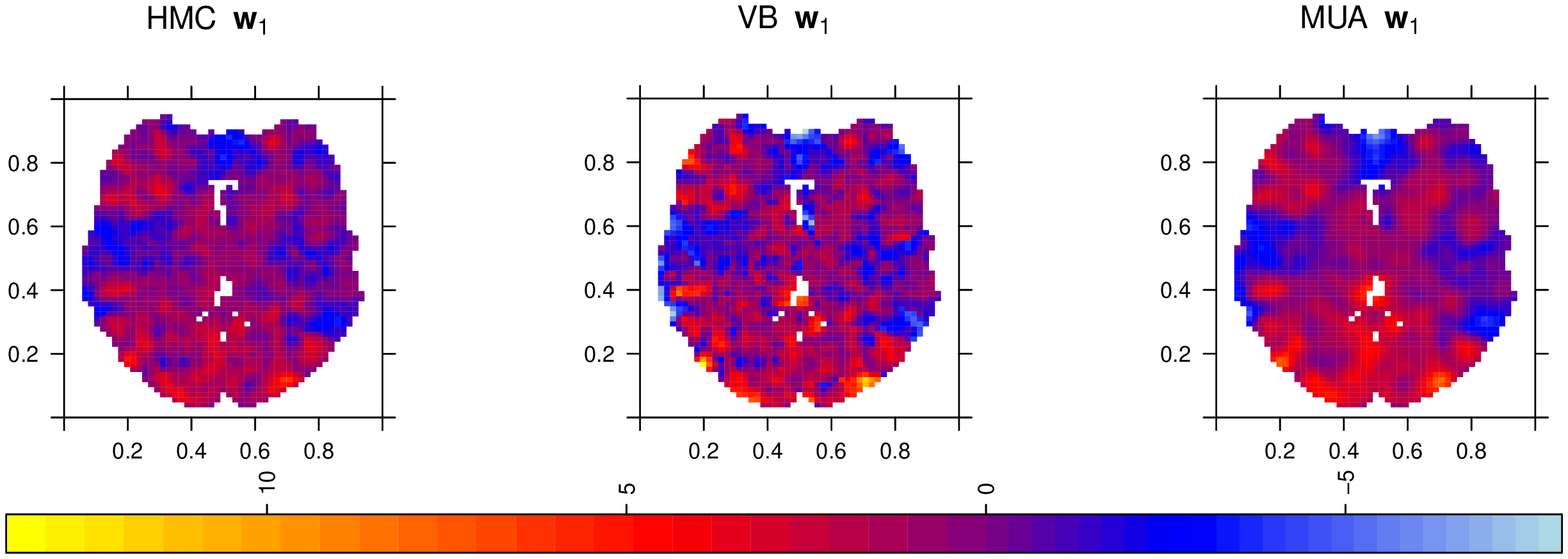}
\caption{\small{\textsf{Posterior mean estimates of $\mathbf w_1$ on the $26^{th}$ plane out of $52$ planes along the z-axis. } }}
\label{fig_real_W_image}
\end{center}
\end{figure}

\begin{figure}[!ht]
\begin{center}
\includegraphics[angle=-90,width=\linewidth]{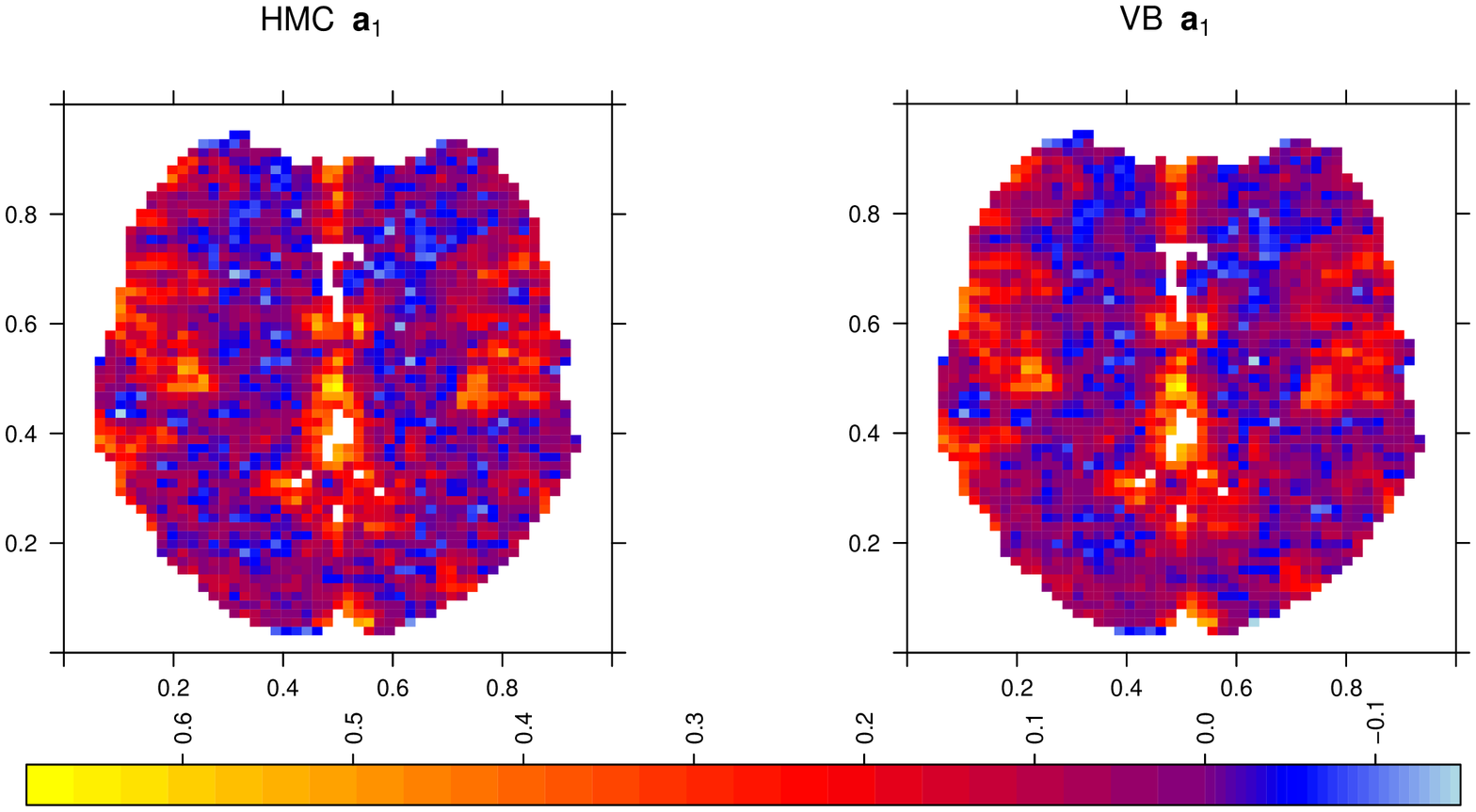}
\caption{\small{\textsf{Posterior mean estimates of $\mathbf a_1$ on the $26^{th}$ plane out of $52$ planes along the z-axis.} }}
\label{fig_real_A_image}
\end{center}
\end{figure}

Images depicting the estimated coefficients are shown in Figures \ref{fig_real_W_image} and \ref{fig_real_A_image}. Due to space restrictions we only display the estimates of $\mathbf w_1$ and $\mathbf a_1$ on the $26^{th}$ plane out of $52$ planes along the z-axis. Additional figures displaying estimates for the other regression coefficients are presented in the  Supplementary Material, Figures 13-14. As seen in the simulation studies, HMC and VB yield very similar posterior mean estimates in terms of auto-regressive coefficients. In terms of the regression coefficients, the estimates from HMC seem to be a bit smoother than those from VB, but still similar in general. Estimates from MUA seem to exhibit a greater degree of spatial smoothing.

\begin{figure}[!ht]
\begin{center}
\includegraphics[angle=-90,width=\linewidth]{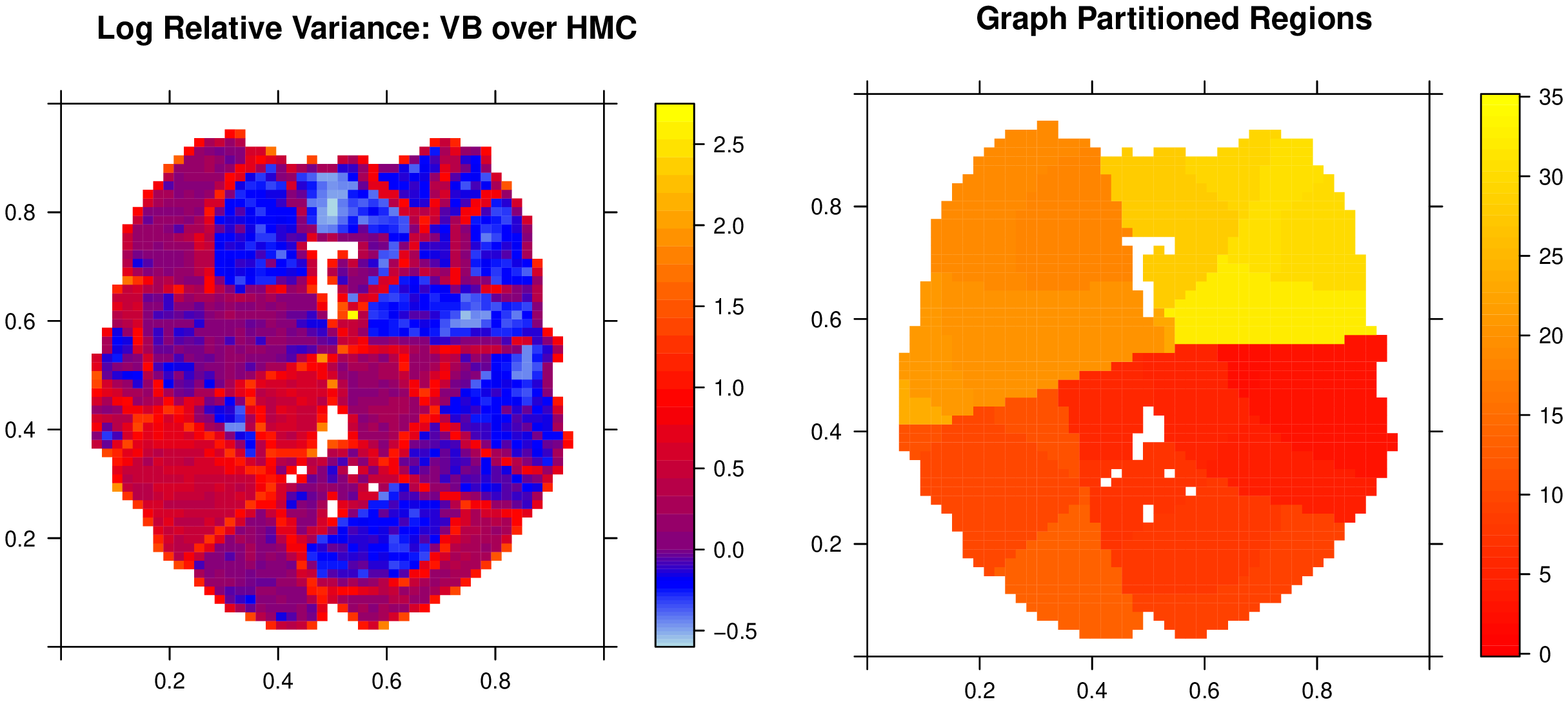}
\caption{\small{\textsf{Log relative ratio of the marginal posterior variance of the regression coefficient obtained from VB over that obtained from HMC. The red-yellow regions in the left image indicate locations where VB results in greater posterior variance relative to HMC for $\mathbf w_1$, the right image shows the graph partitioned regions. Both are from the $26^{th}$ plane out of $52$ planes along the z-axis.} }}
\label{fig_real_var_image}
\end{center}
\end{figure}

To compare VB and HMC with respect to the posterior marginal variance of the regression coefficients, we take the log-ratio of the posterior marginal variance obtained from VB over that obtained from HMC at each voxel, and examine these log-ratio values across all voxels. Doing so we find that for a great proportion of voxels, VB is actually over-estimating the posterior marginal variance relative to HMC. This coincides with the results of our third simulation study. This overestimation may also be arising as a result of the graph partitioning algorithm used in the SPM12 implementation of VB. This is demonstrated in Figure \ref{fig_real_var_image} which depicts an image of the log-ratio marginal-variance values for a single slice for $\mathbf w_1$ alongside the graph partitioned regions, and also in the Supplementary Material, Figure 15, which shows similar images for all of the regression coefficients. From the figures we see that the locations where the posterior marginal variance obtained from VB is higher than that obtained from HMC tend to align with the boundaries of the graph partitioned regions. It appears that the graph partitioning used in the SPM12 implementation contributes to the over-estimation of the posterior variance in this case, as there would be no spatial smoothing across the boundaries of the graph partitioned regions. Practically, this is an important artifact that practitioners should be aware of.

\begin{figure}[!ht]
\begin{center}
\includegraphics[scale=.3]{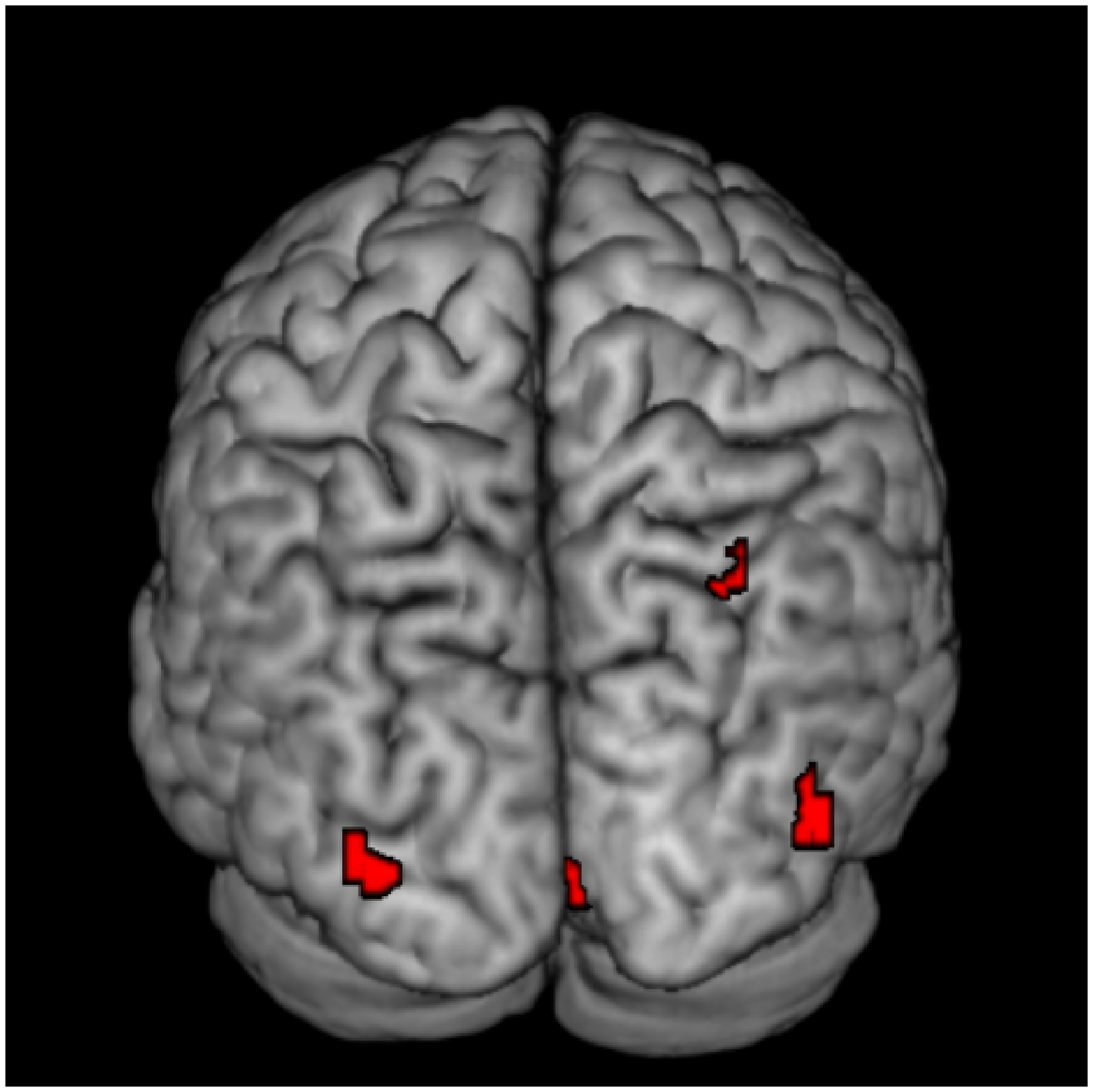}
\includegraphics[scale=.3]{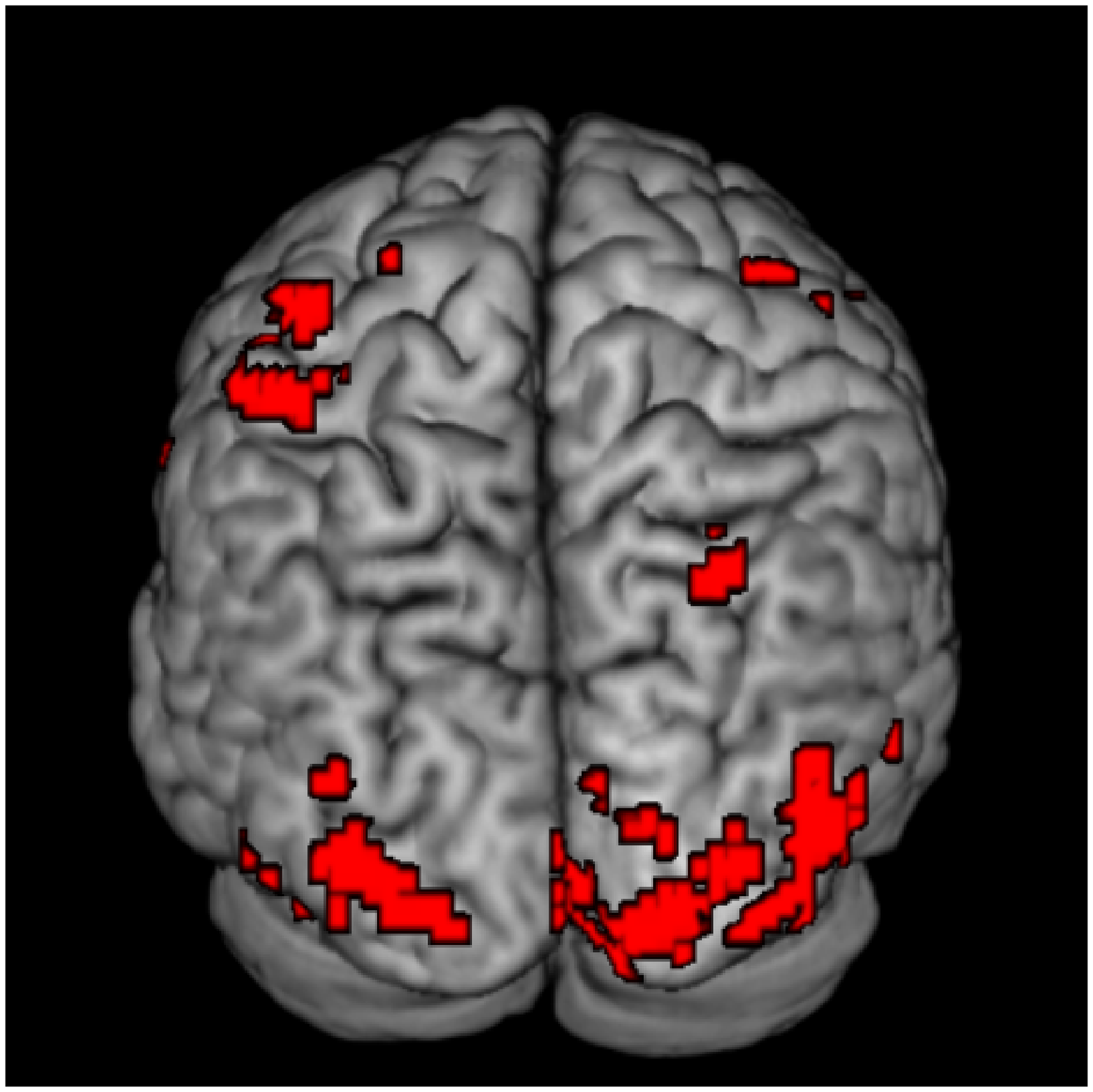}
\caption{\small{\textsf{PPM showing the activated voxels, with an effect size threshold of $1\%$ greater than the global mean and a probability threshold of $95\%$. The left map is obtained from HMC and right map is obtained from VB. The activations are displayed as red dots on a 3-d surface from the posterior view.} }}
\label{fig_ppm}
\end{center}
\end{figure}

We next examine and make comparisons with respect to activations, more specifically, the effect of face by defining a contrast vector $\mathbf c=(1,1,1,1, 0)^T/4$. The value of $\gamma_e$ is set to be $1\%$ greater than the global mean (across voxels) of $\mathbf c^T \mathbf w$ (\cite{ashburner2014spm12}). The value of the probability threshold is set to be $\gamma_p=0.95$. We are then able to compute the PPM and they are depicted in Figure \ref{fig_ppm}. A difference is observed where HMC tends to result in a smaller proportion of the brain indicated as "active" compared with VB. The VB approximation thus results in a wider area of activation when compared with HMC. This result agrees with the simulation results in \cite{yu2016understanding} where the authors suggest that VB seems to report more false positives. Again, this is a practically important difference that practitioners should be aware of.

The PPM's obtained from the two approaches are generally similar, though with more voxels indicated as activated by VB in this particular case. In terms of timing, HMC takes 8.42 hours for 3000 iterations, VB takes 36 minutes, MUA takes 36 seconds with all computations performed on a standard iMac with with 3.2 GHz Intel Core i5.

%% file: paper--discussion.tex
\section{Discussion}

We have compared HMC and mean-field VB for Bayesian inference in the time series analysis of fMRI data with spatial priors. Comparisons were made in three simulation studies with a 2-dimensional grid and an actual single subject fMRI dataset based on a 3-dimensional grid. We have found that for this particular model, under a moderate SNR, HMC and VB provide similar estimates of the posterior distribution, in terms of both point estimation and posterior variability. The VB approximation seems accurate in these cases. However, things do change when one considers a low SNR, where HMC and VB seem to differ substantially in terms of accuracy and inference.

To quantify the relative quality of the approximation and describe how the quality of estimates varies across the three simulation studies we have conducted, we compute the ratio $AMSE_{VB}/AMSE_{HMC}$ separately for each of the $W$ and $A$ parameters in the model, and then average this value across the model parameters to obtain a single measure quantifying the percent increase in the mean-squared-error of estimation obtained when comparing VB to HMC. Doing so, we obtain values of $104$\% in Simulation Study I, $101$\% in Simulation Study II, and $307$\% in Simulation Study III. From this we see clearly that it is in the low SNR setting where the quality of the estimation obtained from VB is lowest relative to HMC. A possible remedy in the framework of VB is to consider more flexible approximations that allow for a greater degree of posterior correlation as in \cite{siden2017fast}. 

In Section 3.3 we found visible differences when comparing the classical and Bayesian approaches. The classical approach does not adopt spatial smoothing priors. Differences seen when comparing HMC to VB in Section 3.3 seem not only due to the mean-field approximation but also due to the graph partitioning algorithm used in the common SPM12 implementation of VB, where VB tends to over-estimate the posterior marginal variance along the edges of the graph partitioned regions. 

In terms of timing, HMC is considerably slower than VB as expected. This is based on running the HMC algorithm for 3000 iterations with the final 1000 iterations used to estimate features of the posterior distribution. We have also run a test case with a much larger Monte Carlo sample of 30000 iterations with the final 15000 iterations used to estimate features of the posterior distribution and have found the results to be very similar to those obtained with the smaller Monte Carlo sample size. As discussed in Neal (2011), $L$ and $\delta$ are parameters that need to be tuned to obtain good performance in the sense that the Markov chain mixes well so that the sampler can adequately traverse the parameter space. For a given value of $L$ our algorithm automatically tunes $\delta$ during the burn-in phase of the HMC algorithm in order to obtain an acceptance rate of approximately 65\%, which is suggested as an approximately optimal acceptance rate by Neal (2011). For choosing $L$ we start with reasonably large values such as $L=100$, so as to avoid random walk behaviour, and increase the value of $L$ based on the output of some preliminary runs of the algorithm. Aside from this approach, there have been some recent developments in this area (see e.g., \cite{hoffman2014no}), but there are still many challenges in tuning and optimizing the HMC algorithm, and in particular for fMRI data where very high-dimensional parameter updates may be required.

Overall, for this particular model and for the settings considered here, our work justifies the use of mean field VB and its implementation in SPM12 under a moderate SNR, while pointing out the disadvantage of VB under a weak SNR, based on our comparisons with the results obtained from HMC. Our work also speaks more generally to the issue of variational Bayes inference and the importance of checking the accuracy of variational Bayes approximations as there is currently no theory that we are aware of guaranteeing the accuracy of these approximations. They are potentially very useful but it is practically important to check their accuracy as we have demonstrated.

%% file: paper--tables.tex
\newpage

\section*{Tables}

\begin{table}[!ht]
\begin{center}
\begin{tabular}{cccccccc}
\hline \hline
methods	&	measure	&	W1	&	W2	&	W3	&	W4	&	W5	&	A1	\\
true	&	Moran's I	&	0.121	&	0.169	&	0.136	&	0.187	&	0.122	&	0.179	\\ \hline
HMC	&	ASBIAS	&	0.123	&	0.120	&	0.105	&	0.110	&	0.001	&	4.52E-04	\\
	&	AMSE	&	0.405	&	0.452	&	0.412	&	0.420	&	0.007	&	1.19E-03	\\
	&	AVAR	&	0.411	&	0.468	&	0.425	&	0.435	&	0.008	&		\\
	&	Correlation	&	0.997	&	0.999	&	0.998	&	0.998	&	1.000	&	0.995	\\
	&	Moran's I	&	0.123	&	0.171	&	0.137	&	0.189	&	0.122	&	0.182	\\ \hline
VB	&	ASBIAS	&	0.082	&	0.070	&	0.068	&	0.079	&	0.001	&	4.11E-04	\\
	&	AMSE	&	0.433	&	0.470	&	0.424	&	0.441	&	0.007	&	1.24E-03	\\
	&	AVAR	&	0.460	&	0.510	&	0.459	&	0.470	&	0.008	&		\\
	&	Correlation	&	0.997	&	0.999	&	0.998	&	0.998	&	1.000	&	0.995	\\
	&	Moran's I	&	0.123	&	0.171	&	0.137	&	0.189	&	0.122	&	0.186	\\
\hline \hline
\end{tabular}
\end{center}
\caption{\small{\textsf{Summary statistics for Simulation Study I. The results from VB are presented as a percentage of those obtained HMC. The true value of Moran's I is listed for each regressor in the first row as a reference.}}}
\label{tab_K5P1}
\end{table}

\begin{table}[!ht]
\begin{center}
\footnotesize{
\begin{tabular}{cccccccccc}
\hline \hline
methods	&	measure	&	W1	&	W2	&	W3	&	W4	&	W5	&	W6	&	W7	&	W8	\\
true	&	Moran's I	&	0.111	&	0.137	&	0.151	&	0.144	&	0.125	&	0.128	&	0.121	&	0.109	\\ \hline
HMC	&	ASBIAS	&	0.054	&	1.097	&	0.711	&	0.115	&	0.973	&	0.840	&	0.108	&	0.860	\\
	&	AMSE	&	0.610	&	4.336	&	3.566	&	0.549	&	2.317	&	2.160	&	0.444	&	1.840	\\
	&	AVAR	&	0.617	&	4.181	&	3.466	&	0.561	&	2.244	&	2.112	&	0.459	&	1.807	\\
	&	Correlation	&	1.000	&	0.998	&	0.999	&	0.999	&	0.991	&	0.992	&	0.998	&	0.982	\\
	&	Moran's I	&	0.111	&	0.139	&	0.152	&	0.145	&	0.128	&	0.132	&	0.123	&	0.112	\\ \hline
VB	&	ASBIAS	&	0.039	&	0.669	&	0.405	&	0.072	&	0.866	&	0.756	&	0.084	&	0.877	\\
	&	AMSE	&	0.622	&	4.379	&	3.566	&	0.571	&	2.271	&	2.117	&	0.453	&	1.785	\\
	&	AVAR	&	0.623	&	4.765	&	3.813	&	0.595	&	2.401	&	2.218	&	0.477	&	1.807	\\
	&	Correlation	&	1.000	&	0.998	&	0.999	&	0.999	&	0.991	&	0.992	&	0.998	&	0.982	\\
	&	Moran's I	&	0.111	&	0.139	&	0.152	&	0.145	&	0.131	&	0.135	&	0.124	&	0.119	\\ \hline \hline
	&		&	W9	&	W10	&	W11	&	W12	&	W13	&	A1	&	A2	&	A3	\\
true	&	Moran's I	&	0.104	&	0.148	&	0.189	&	0.130	&	0.128	&	0.108	&	0.127	&	0.174	\\ \hline
HMC	&	ASBIAS	&	0.761	&	0.123	&	0.639	&	0.576	&	0.002	&	4.71E-04	&	3.72E-04	&	3.13E-04	\\
	&	AMSE	&	1.639	&	0.369	&	1.197	&	1.203	&	0.009	&	1.19E-03	&	8.95E-04	&	5.58E-04	\\
	&	AVAR	&	1.607	&	0.384	&	1.126	&	1.211	&	0.009	&		&		&		\\
	&	Correlation	&	0.980	&	0.996	&	0.977	&	0.983	&	1.000	&	0.992	&	0.988	&	0.975	\\
	&	Moran's I	&	0.108	&	0.151	&	0.198	&	0.133	&	0.128	&	0.111	&	0.129	&	0.182	\\ \hline
VB	&	ASBIAS	&	0.883	&	0.108	&	0.658	&	0.541	&	0.002	&	4.52E-04	&	4.58E-04	&	3.10E-04	\\
	&	AMSE	&	1.672	&	0.380	&	1.221	&	1.155	&	0.009	&	1.25E-03	&	9.04E-04	&	5.47E-04	\\
	&	AVAR	&	1.559	&	0.399	&	1.227	&	1.260	&	0.009	&		&		&		\\
	&	Correlation	&	0.980	&	0.996	&	0.977	&	0.983	&	1.000	&	0.992	&	0.988	&	0.975	\\
	&	Moran's I	&	0.117	&	0.153	&	0.212	&	0.138	&	0.128	&	0.113	&	0.135	&	0.191	\\
\hline \hline
\end{tabular}}
\end{center}
\caption{\footnotesize{\textsf{Summary statistics for Simulation Study II. The results from VB are presented as a percentage of those obtained HMC. The true value of Moran's I is listed for each regressor in the first row as a reference.}}}
\label{tab_K13P3}
\end{table}

\begin{table}[!ht]
\begin{center}
\begin{tabular}{cccccccc}
\hline \hline
methods	&	measure	&	W1	&	W2	&	W3	&	W4	&	W5	&	A1	\\
true	&	Moran's I	&	0.167	&	0.203	&	0.202	&	0.197	&	0.147	&	0.122	\\ \hline
HMC	&	ASBIAS	&	0.100	&	0.089	&	0.098	&	0.081	&	0.001	&	5.44E-04	\\
	&	AMSE	&	0.163	&	0.168	&	0.175	&	0.165	&	0.046	&	1.56E-03	\\
	&	AVAR	&	0.121	&	0.149	&	0.143	&	0.163	&	0.045	&		\\
	&	Correlation	&	0.916	&	0.948	&	0.942	&	0.961	&	1.000	&	0.997	\\
	&	Moran's I	&	0.201	&	0.260	&	0.193	&	0.209	&	0.147	&	0.124	\\ \hline
VB	&	ASBIAS	&	0.019	&	0.016	&	0.019	&	0.017	&	0.001	&	4.52E-04	\\
	&	AMSE	&	0.678	&	0.734	&	0.648	&	0.677	&	0.051	&	1.58E-03	\\
	&	AVAR	&	0.534	&	0.583	&	0.503	&	0.535	&	0.045	&		\\
	&	Correlation	&	0.980	&	0.986	&	0.990	&	0.990	&	1.000	&	0.997	\\
	&	Moran's I	&	0.103	&	0.143	&	0.104	&	0.130	&	0.147	&	0.124	\\
\hline \hline
\end{tabular}
\end{center}
\caption{\small{\textsf{Summary statistics for Simulation Study III. The results from VB are presented as a percentage of those obtained HMC. The true value of Moran's I is listed for each regressor in the first row as a reference.}}}
\label{tab_SNR1e-3}
\end{table}

\begin{table}[!ht]
\begin{center}
\begin{tabular}{ccccccc}
\hline \hline
											
Correlation	&	$\mathbf w_1$	&	$\mathbf w_2$	&	$\mathbf w_3$	&	$\mathbf w_4$	&	$\mathbf w_5$	\\ \hline
(VB, HMC)	&	0.91	&	0.93	&	0.92	&	0.91	&	1.00	\\
(MUA, HMC)	&	0.87	&	0.84	&	0.84	&	0.83	&	0.66	\\
\hline \hline
\end{tabular}
\end{center}
\caption{\small{\textsf{Correlation (across voxels) in the estimated regression coefficients obtained from HMC and VB, and HMC and MUA.}}}
\label{tab_corr}
\end{table}